\documentclass[10pt,a4paper,final]{iopartMod} 

\expandafter\let\csname equation*\endcsname\relax 
 \expandafter\let\csname endequation*\endcsname\relax 
\usepackage{amsmath,amssymb}
\usepackage{bm}
\usepackage{graphicx,xcolor}
\usepackage[colorlinks=true,citecolor=blue]{hyperref}
\usepackage{color,graphicx}

\newcommand{ \calA }{ \mathcal{A} }

\renewcommand{\Re}{\operatorname{Re}}
\renewcommand{\Im}{\operatorname{Im}}

\begin{document}

\title[Gravitational lensing beyond geometric optics: \textsc{I}]{Gravitational lensing beyond geometric optics: I. Formalism and observables}
\author{Abraham I. Harte}
\address{Centre for Astrophysics and Relativity, School of Mathematical Sciences
	\\
	Dublin City University, Glasnevin, Dublin 9, Ireland}

\ead{abraham.harte@dcu.ie}

\begin{abstract}
The laws of geometric optics and their corrections are derived for scalar, electromagnetic, and gravitational waves propagating in generic curved spacetimes. Local peeling-type results are obtained, where different components of high-frequency fields are shown to scale with different powers of their frequencies. Additionally, finite-frequency corrections are identified for a number of conservation laws and observables. Among these observables are a field's energy and momentum densities, as well as several candidates for its corrected ``propagation directions.'' 
\end{abstract}

\section{Introduction}

Nearly all astronomical observations involve, fundamentally, measurements of electromagnetic or (more recently) gravitational radiation. However, these waves carry with them an imprint of the spacetime through which they travel. The spacetime geometry provides a kind of ``transfer function'' that relates the intrinsic properties of a source to its radiated fields. Such relations must be understood if an object's properties are to be accurately inferred from distant measurements of its fields. If a source has already been characterized, its radiation might instead be used to probe the intervening geometry, and thus the matter which contributes to it---matter which might not be bright enough to observe directly. For these reasons and others, gravitational lensing has become a standard tool with which to extract information from astronomical observations. 

Much of the theory of gravitational lensing which is used in practice may be viewed as an elaboration on the particle-like laws of geometric optics: Light travels along null geodesics, intensity variations are determined by the changing cross-sectional areas of ray bundles, and polarization vectors are parallel transported. These simple statements beget a remarkable variety of applications \cite{SchneiderEhlers,WambsganssRev, Perlick, BartelmannRev}. However, the laws of geometric optics are an approximation. Electromagnetic fields are more properly described as solutions to Maxwell's equations, and gravitational waves as solutions to Einstein's equation. While the full complexities of these equations may often be ignored, there are exceptions. For example, it is well-known in ordinary optics \cite{CausticsBook, BornWolf, ThorneBlandford} that the geometric approximation breaks down completely at caustics---a result which has also had astrophysical implications \cite{SchneiderEhlers, BlandfordCaustics, Turyshev2017a}. In other contexts, wave-optical corrections may be small but still detectable, in which case they might supply information which is different---and therefore complementary to---that which can be learnt from geometric optics alone.

Wave-optical effects may be viewed as frequency-dependent corrections to the frequency-independent laws of geometrical optics. Apparent source locations, intensities, phases, and polarization states might all depend on the frequencies at which a source is observed. Any such quantity measured at a sufficiently-high characteristic frequency $\omega$ may be viewed as a geometric-optics result plus relative corrections which scale like, e.g., $\omega^{-1}$.  Somewhat more precisely, these corrections scale like $(\omega \ell)^{-1}$, where $\ell$ is a relevant lengthscale. Several lengthscales may be present simultaneously and different ones can be relevant for different observables. In simple cases, $\ell$ might represent a notion of distance between a source and its observer: That geometric optics breaks down at caustics may be understood in this context by noting that the ``source-centered area distance'' $r_\mathrm{a}$ goes to zero at caustics and $\ell \sim r_\mathrm{a}$ for some contributions to some observables. More generally,  $\ell$ can be a nontrivial composite of different lengthscales. For example, some corrections associated with fields of  mass $\mu \neq 0$ which are radiated by a source at affine distance $r$ can involve the lengthscale $\ell \sim (\mu^2 r)^{-1}$; fractional corrections to geometric optics grow with distance for massive fields. More generally (and even for massless fields) a relevant $\ell$ might be a highly nontrivial nonlocal combination of different lengthscales---including those associated with the spacetime geometry and with details of the particular field under consideration. A systematic development of the underlying theory is thus required in order to understand precisely when such effects might be interesting. This paper begins on the path to such a development. 

More directly, the purpose here is to provide general expressions which allow the propagation of high-frequency scalar, electromagnetic, and gravitational waves to be characterized in general spacetimes. While the basic equations governing geometric optics and its corrections have been discussed elsewhere \cite{EhlersGeoOptics, Anile, Isaacson1} from a general spacetime perspective, very few of their higher-order consequences appear to have been explored. Some discussions which do go beyond geometric optics have appeared in various contexts, although most of these have employed a different ``pseudo-Euclidean'' approach which is restricted to weakly-curved spacetimes \cite{SchneiderEhlers, Nakamura1998, Nakamura1999, Takahashi2006, gravGrating}. 

The discussion here is intended to be largely self-contained, and therefore begins by reviewing the equations which govern geometric optics and its corrections. Mathematically, these equations transform the partial differential equations which control the underlying fields into a hierarchy of algebraic constraints and ordinary differential equations along null geodesics. These are used to derive wave-optical corrections to field strengths, curvature perturbations, stress-energy tensors, and conservation laws---in arbitrary spacetimes and for arbitrary polarization states. Several types of ``propagation direction'' are identified and discussed. For some such definitions, multiple directions can arise simultaneously; these experience relative corrections which scale like $\omega^{-1/2}$ instead of, e.g., $\omega^{-1}$, implying that they are particularly sensitive to wave-optical effects. Frequency dependencies of the different tensorial components of electromagnetic and gravitational waves are determined as well, resulting in what are essentially local peeling results. Throughout, we emphasize connections between the various types of fields considered here. When, for example, can aspects of an electromagnetic problem be reduced to those of an effective scalar problem?

\noindent
\textit{Notation}---Sign and index conventions follow those of Wald \cite{Wald}. Units are used in which $G=c=1$ and the number of spacetime dimensions is fixed at four. In several cases, a complex field is considered despite that it is only its real component which is considered to be physical. These fields are distinguishing by using an upper-case symbol to denote the real quantity and a lower-case one for its complex counterpart; $F_{ab} = \Re f_{ab}$, for example. 

\section{Scalar fields}
\label{Sect:Scalar}

Derivations of geometric optics and its corrections are typically approached  at the level of freely-propagating fields, without considering how those fields are produced. This is also the perspective adopted here. The general method is simplest to understand for a freely-propagating, real scalar field $\Psi$, which is the first case we consider. Suppose in particular that this field satisfies the source-free Klein-Gordon equation 
\begin{equation}
	(\Box - \xi R - \mu^2) \Psi = 0
	\label{KleinGordon}
\end{equation}
on a fixed background spacetime $(M,g_{ab})$, where $R$ denotes the Ricci scalar associated with this background, $\Box \equiv \nabla^a \nabla_a$, and the field mass $\mu$ and the curvature coupling $\xi$ are constants. Approximate solutions may be found by restricting the geometry, the initial data for $\Psi$, or the spacetime region of interest. Here, we place no significant restrictions on the geometry, nor do we require that the field be evaluated in any special location. Instead, we restrict the initial data in the sense of imposing a high-frequency ansatz. The associated approximation is systematic in the sense that geometric optics is recovered as the first term in an easily-derived perturbative expansion. While there are systems in which the laws of geometric optics arise without any significant frequency restrictions \cite{Teitelboim, HoganEllis, NolanGeo}, these are largely special cases wherein no relevant lengthscale exists which might be used to decide whether a particular frequency is large or small.

Physically, the connection between high frequencies and geometric optics may be understood by noting that discontinuities in the field---perhaps jumps representing bits of information transmitted from a source to a waiting receiver---may be expected to obey geometrical laws. The essential structure of these discontinuities is however determined by shorter wavelengths than any scales which might be associated either with the background geometry or the curvature of a wavefront; geometric optics should thus be recovered at high frequencies. To motivate that high-frequency assumptions are not only sufficient but also ``not too strong,'' recall that Huygens' principle is valid essentially just for massless fields propagating in very particular spacetimes \cite{HuygensBook, HuygensRev}: Except in special cases, finite-frequency data is known to propagate in timelike as well as null directions---a process which cannot be described by the geometric-optics expectation that information travels only along null geodesics. The geometric picture must therefore fail  unless interference can be counted upon to suppress propagation in timelike directions. This type of suppression is exactly what occurs at high frequencies.

Applying a high-frequency approximation now requires that we say what exactly is meant by ``frequency.'' The concept is not \textit{a priori} well-defined without reference to an observer, and no observer naturally presents itself (except in special spacetimes). We proceed instead by applying a WKB ansatz, where the frequency is simply identified with an expansion parameter $\omega$. More precisely, consider a 1-parameter family of real solutions $\Psi(x;\omega)$ to the Klein-Gordon equation which can be expanded asymptotically as real components of a complex series with the form\footnote{Various other WKB-like ans\"{a}tze may be considered. For example, the amplitude might be replaced with a single $\omega$-independent function while the exponent is instead expanded in powers of $\omega^{-1}$. This naturally leads to the consideration of exponents which are not necessarily purely imaginary, thus allowing evanescent waves and other exponentially-suppressed phenomena to potentially be understood. Such an ansatz nevertheless comes with considerable complications, and is not considered here. These complications are especially severe when considering electromagnetic or gravitational fields, in which case one must resort to ``phases'' which take the form of higher-rank tensor fields.}
\begin{equation}
  \psi(x;\omega) = e^{i \omega \varphi(x)} \sum_{n = 0}^\infty \omega^{-n} \calA_n (x) .
  \label{PhiAnsatz}
\end{equation}
Here, the phase function $\varphi(x)$ is real and the amplitudes $\calA_n$ may be complex. Physically, this expresses the intuitive concept of a locally-planar field with real, non-constant phase $\omega \varphi(x)$. The parameter $\omega$ scales that phase and also the frequencies associated with any particular observer which might exist. The remainder of this paper refers to the limit $\omega \to \infty$ as the geometric-optics limit. As is usual for asymptotic series, the infinite upper limit in \eqref{PhiAnsatz} is formal. While the series does not necessarily converge, there is a sense in which finite truncations at order $m$ may be arranged to satisfy the Klein-Gordon equation up to terms of order $\omega^{-m}$ as $\omega \to \infty$. This result is obtained by substituting the ansatz for $\psi$ into the field equation \eqref{KleinGordon} and equating equal powers of $\omega$, a method which appears first to have been introduced in an optical context by Sommerfeld and Runge \cite{SommerfeldRunge}. We now apply it for Klein-Gordon fields in general spacetimes.

\subsection{Geometric optics} 

Assuming that $\calA_0 \neq 0$, the leading-order consequence of substituting \eqref{PhiAnsatz} into \eqref{KleinGordon} is the well-known eikonal equation
\begin{equation}
  k^a k_a = 0, \qquad k_a \equiv - \nabla_a \varphi,
  \label{eikonal}
\end{equation}
which implies that hypersurfaces of constant $\varphi$ must be null. It follows directly from this that $\nabla_{[a} k_{b]} = \nabla_{[b} \nabla_{a]} \varphi = 0$ and $(k \cdot \nabla) k_a = 0$. The integral curves of $k^a$ thus form a twist-free null geodesic congruence. They are the rays of geometric optics. In more mathematical language, the hypersurfaces of constant $\varphi$  are characteristics of the Klein-Gordon equation. Similarly, the rays are bicharacteristics; see, e.g., \cite{Friedlander} for definitions of these terms.

Applying the field equation to one higher order constrains the zeroth-order amplitude $\calA_0$ via
\begin{equation}
	L \calA_0 = 0,
	  \label{phi0trans}
\end{equation}
where 
\begin{equation}
	L \equiv 2 k \cdot \nabla + (\nabla \cdot k) 
	\label{transOp}
\end{equation}
is a transport operator associated with the given ray system. This $L$ may be viewed as an \textit{ordinary} differential operator along each null ray tangent to $k^a$, implying that \eqref{phi0trans} may be treated as an ordinary differential equation---or transport equation---for $\calA_0$ along the rays. Amplitudes evaluated on distinct rays thus propagate independently of one another; cross-ray interaction does not exist at this order. The field mass $\mu$ and the curvature coupling $\xi$ are also irrelevant at this order.

An interpretation for the leading-order transport equation may be gained by using it to show that
\begin{equation}
	J^a_0 \equiv |\calA_0|^2 k^a
	\label{J0scalar}
\end{equation}
is a conserved current, which makes precise a sense in which $|\calA_0|^2 \times \mbox{(cross-sectional area of beam})$ is constant along each ray. Also noting that $|\calA_0|^2$ is shown below to determine the scale of the leading-order energy density associated with $\Psi = \Re \psi$, the area-intensity law of geometric optics is seen to be encoded in the transport equation for $\calA_0$.

General trends in the evolution of $|\calA_0|^2$ may be understood in somewhat more detail by recalling the Raychaudhuri equation in the form
\begin{equation}
	k \cdot \nabla (\nabla \cdot k) = -(R_{ab} k^a k^b + \nabla^a k^b \nabla_a k_b).
\end{equation}
The right-hand side here cannot be positive if the null energy condition is satisfied for the Ricci tensor $R_{ab}$, meaning that it is impossible in these cases for $\nabla \cdot k$ to increase along rays. Hence,
\begin{equation}
	(k \cdot \nabla)^2 \ln |\calA_0|^2 \geq 0,
\end{equation}
which suggests that intensities tend to increase along rays, at least eventually. This is taken to an extreme at caustics, where $|\calA_0|^2 \to \infty$ and $\nabla \cdot k \to -\infty$. 

Returning to the overall interpretation of \eqref{phi0trans}, it is not only the magnitude of $\calA_0$ which is physically significant. Its complex argument may be important as well, and the transport equation implies that this must be constant along rays:
\begin{equation}
	k \cdot \nabla \arg \calA_0 = 0.
	\label{noPhase}
\end{equation}
The meaning of this may be understood by first observing that there is a degeneracy in writing the leading-order field in the form $\calA_0 e^{i \omega \varphi}$; it may also be written as
\begin{equation}
	\calA_0 e^{i \omega \varphi} = |\calA_0| \exp \big[ i\omega (\varphi + \omega^{-1} \arg \calA_0)\big],
	\label{phaseGuess}
\end{equation}
which suggests that 
\begin{equation}
	\hat{\varphi} \equiv \varphi + \omega^{-1} \arg \calA_0
	\label{phaseCorrect}
\end{equation}
might be viewed as a corrected phase function. Recalling \eqref{eikonal}, this also suggests the corrected propagation direction
\begin{equation}
	\hat{k}_a \equiv - \nabla_a \hat{\varphi} = k_a - \omega^{-1} \nabla_a \arg \calA_0.
	\label{Kscalar}
\end{equation}
The constant-phase result \eqref{noPhase} implies that this is null and geodesic to the given order:
\begin{equation}
	\hat{k} \cdot \nabla \hat{k}_a = \mathcal{O}(\omega^{-2}), \qquad \hat{k} \cdot \hat{k} = \mathcal{O}(\omega^{-2}).
\end{equation}
Eq. \eqref{TScalarFact} below confirms that $\hat{k}_a$ does indeed describe the direction of the field's 4-momentum density at leading and subleading orders, as seen by any observer. It is our first post-geometric optics correction.

\subsection{Corrections to geometric optics}

More general corrections to geometric optics arise from the higher-order amplitudes $\calA_n$ which appear in the expansion \eqref{PhiAnsatz}. These are constrained by considering higher powers of $\omega^{-1}$ which arise when substituting that expansion into the field equation. To all orders, this procedure results in transport equations which act only along the null geodesic rays of the leading-order geometric optics solution; the corrected propagation vector given by \eqref{Kscalar} never arises in this way. In fact, all higher-order transport equations involve the same transport operator \eqref{transOp} which appears at zeroth order: For all $n \geq 1$,
\begin{equation}
	L \calA_n = -i (\Box - \xi R - \mu^2)  \calA_{n-1}.
	\label{phiNtrans}
\end{equation}
That the same $L$ arises for all $n$ implies that it is impossible for the aforementioned failure of Huygens' principle to ever be taken into account by the WKB ansatz. So-called tail effects, which involve the propagation of fields in timelike directions, thus fail to be taken into account not only by geometric optics, but also by all of its corrections in integer powers of $\omega^{-1}$. This is mathematically consistent in the sense that the expansion is intended only to be an asymptotic approximation; it cannot be used to describe effects which are, e.g., exponentially suppressed as $\omega \to \infty$. Tails are examples of such effects. In this context, they are intrinsically non-perturbative.

Despite this limitation, much can be learned by examining the higher-order terms in \eqref{PhiAnsatz}. If the right-hand side of \eqref{phiNtrans} is nonzero and if $\calA_0 \neq 0$ everywhere of interest, the operator identity $L \calA_0 = 2 \calA_0 k \cdot \nabla$ implies that
\begin{equation}
	k \cdot \nabla (\calA_n/\calA_0) = - \frac{i}{2} \calA_0^{-1} ( \Box - \xi R - \mu^2 ) \calA_{n-1}.
	\label{phiNtrans2}
\end{equation}
Any change in $\calA_n/\calA_0$ along a ray segment is therefore given by the integral of the right-hand side of this equation along that segment, where the integral is performed along a ray with respect to an ``affine radius'' $r(x)$ for which
\begin{equation}
	k \cdot \nabla r =1.
\label{rDef}
\end{equation}
The appearance of the $\Box$ operator in \eqref{phiNtrans2} effectively allows for interference between neighboring rays. More generally, the right-hand side measures the extent by which $\calA_{n-1}$ fails to satisfy the field equation. If any $\calA_m$ did satisfy that equation exactly, the expansion for $\psi$ would exactly terminate at order $m$ (ignoring homogeneous solutions which can always be added to the $\calA_n$ if no initial conditions are imposed).

In summary, asymptotic approximations for real high-frequency solutions $\Psi = \Re \psi$ of the Klein-Gordon equation may be generated by combining the ansatz \eqref{PhiAnsatz} for a complex $\psi$ with the eikonal equation \eqref{eikonal} and the transport equations \eqref{phi0trans} and \eqref{phiNtrans} [or \eqref{phiNtrans2}]. These results convert the partial differential equation which governs $\Psi$ into a collection of ordinary differential equations for the $\calA_n$. Similar equations have been been obtained before for ordinary optics in flat spacetime and in the presence of nontrivial materials \cite{BornWolf, SommerfeldRunge, Keller, Kline}, and also for electromagnetic and gravitational waves propagating in vacuum in generic background spacetimes \cite{ThorneBlandford, EhlersGeoOptics, Anile, Isaacson1, MTW}. 

\subsection{Observables}
\label{Sect:ScalarObs}

We now apply the high-frequency ansatz and the associated transport equations in order to compute various observables which depend on $\psi$, or its real component $\Psi$. In many cases, one does not measure these fields themselves, but rather their gradients---which may be inferred by, e.g., observing the motions of test charges. Using \eqref{PhiAnsatz} and \eqref{eikonal} while defining $\calA_{-1} \equiv 0$, this gradient admits the expansion
\begin{equation}
	\nabla_a \psi = -i \omega e^{i \omega \varphi} \sum_{n=0}^\infty \omega^{-n} \left( \calA_n k_a + i \nabla_a \calA_{n-1} \right).  
	\label{gradPsi}
\end{equation}
Test particles thus experience forces\footnote{The force acting on a scalar test particle with charge $q$ which is immersed in a (real) field $\Psi$ is known to be $q \nabla_a \Psi$, where the notion of force used here refers to the rate of change of a particle's momentum. Note that force is sometimes defined instead in terms of a particle's rest mass multiplied by its 4-acceleration, which can be different. Regardless, the expression here may be motivated using the actions discussed in, e.g., \cite{Quinn, HartePoisson}. It may also be derived by appending a source term to the wave equation \eqref{KleinGordon} and then applying stress-energy conservation using \eqref{Tscalar}; see \cite{HarteReview} for a full discussion of the minimally-coupled case, also including extended-body and self-interaction effects.} parallel to $k^a$ at leading nontrivial order. Noting that $k^a$ is null and the trajectory of a massive test particle must be timelike, the leading-order force inevitably changes a particle's rest mass while also accelerating it along (or against) the direction of propagation. Forces transverse to $k^a$ may appear at higher orders. For example, if $\calA_0 \neq 0$,
\begin{equation}
	\nabla_a \psi = - i \omega |\calA_0| e^{i\omega \hat{\varphi}} \left[ 1 + \omega^{-1}( \calA_1/\calA_0)\right] \left[ \hat{k}_a + i \omega^{-1} \nabla_a \ln |\calA_0| + \mathcal{O}(\omega^{-2}) \right], 
	\label{scalarForce}
\end{equation}
where $\hat{\varphi}$ is the corrected phase \eqref{phaseCorrect} and $\hat{k}_a$ the associated direction \eqref{Kscalar}. Although $\hat{k}_a$ is  identified above as a corrected propagation direction, it does not necessarily determine the direction of the force at this order. However, that portion of the force which does not lie along $\hat{k}_a$ is completely out of phase with that portion which does.

At least in electromagnetic applications, the effects of high-frequency fields are not typically observed by tracking their effects on individual test particles. Instead, measurements are often taken of quantities which depend on suitably-defined averages of a field's stress-energy tensor. Averaging may be intended in different contexts as being performed over time, space, or ensemble. Instead of entering into such distinctions here, consider a general observable $S[\psi]$ which is homogeneous and quadratic in $\psi$, in the sense that it can be written as $\hat{S}[\psi,\psi]$, where
\begin{equation}
	\hat{S}[\psi,\psi'] \equiv \frac{1}{4} \big( S[ \psi + \psi' ] - S[\psi - \psi'] \big) 
	\label{Sbilinear}
\end{equation}
is bilinear and symmetric. In terms of this, we define ``the'' average of $S[\Psi]$ to be
\begin{equation}
	\langle S \rangle \equiv \frac{1}{2} \hat{S} [ \psi, \bar{\psi} ] 
	\label{avDef}
\end{equation}
for any field $\Psi = \Re \psi$ which may be expanded via the WKB ansatz \eqref{PhiAnsatz}. This is equivalent to simply removing all terms in $S[\Psi]$ which depend explicitly on $e^{\pm 2 i \omega \varphi}$.

Perhaps the simplest application of this prescription is to the square of the field itself, i.e. the functional $S[\psi] = \psi^2$; employing \eqref{avDef},
\begin{align}
	\langle \Psi^2 \rangle \equiv \frac{1}{2} |\psi|^2 = \frac{1}{2} |\calA_0 + \omega^{-1} \calA_1|^2 + \mathcal{O}(\omega^{-2}).
	\label{Psi2Av}
\end{align}
It is shown below that this and $\hat{k}_a$ control the scale of the averaged stress-energy tensor, and thus, e.g., observed energy and momentum densities; cf. \eqref{omegaO} and \eqref{Iscalar}. If the complex phase of $\calA_0$ is trivial in the sense that $\nabla_a \arg \calA_0 = 0$, the transport equation \eqref{phiNtrans} implies that (ignoring homogeneous solutions) $\calA_1$ is out of phase with $\calA_0$ and $|\calA_0 + \omega^{-1} \calA_1|^2 = |\calA_0|^2 + \mathcal{O}(\omega^{-2})$. Nontrivial intensity corrections at subleading order thus require that $\nabla_a \arg \calA_0 \neq 0$.

Another quadratic observable which may be considered is $\langle |\nabla \Psi|^2 \rangle$. Using the transport equations together with \eqref{avDef} to again define the average, its first nonzero term is found to be
\begin{align}
	\langle |\nabla \Psi|^2 \rangle = \frac{1}{2} \left( |\nabla \calA_0|^2 +  \Re \left[ \calA_0 (\Box - \xi R - \mu^2) \bar{\calA}_0 \right] \right) + \mathcal{O}(\omega^{-1}).
	\label{Dpsi2}
\end{align}
Although this is suppressed by two powers of $\omega$ with respect to generic components of $\langle \nabla_a \Psi \nabla_b \Psi \rangle = \mathcal{O}(\omega^2)$, it is locally determined only by the leading-order amplitude $\calA_0$. This characteristic---where results beyond geometric optics are locally and completely determined only by quantities which are already well-defined in geometric optics---is shared by many of the results below. It may also be noted that $\langle |\nabla \Psi|^2 \rangle$ depends on $\xi$ and $\mu$ only via terms which measure the extent by which $\calA_0$ fails to satisfy the Klein-Gordon equation.

The most important quadratic observable associated with $\Psi$ is perhaps its stress-energy tensor. This may be defined using the functional derivative of the action with respect to $g^{ab}$ \cite{Wald}, and recalling that the Klein-Gordon equation \eqref{KleinGordon} follows from the Lagrangian $- 2 [ \nabla_a \Psi \nabla^a \Psi + (\mu^2 + \xi R) \Psi^2]$, the stress-energy tensor associated with $\psi$ must be\footnote{The factor of $-2$ in the Lagrangian quoted here is a matter of convention. Changing it would modify the factor of $1/4\pi$ in \eqref{Tscalar}.}  \cite{FlanaganWald}
\begin{align}
  T_{ab}[\psi] = \frac{1}{4\pi} \Big\{ \nabla_a \psi \nabla_b \psi - \tfrac{1}{2} g_{ab} \left(\nabla^c \psi \nabla_c \psi + \mu^2 \psi^2 \right) + \xi \big[ (R_{ab} - \tfrac{1}{2} g_{ab} R ) \psi^2 
  \nonumber
  \\
  ~ - 2 (\delta^c_a \delta^d_b - g_{ab} g^{cd}) \nabla_c (\psi \nabla_d \psi) \big] \Big\}.
  \label{Tscalar}
\end{align}
It follows from \eqref{Sbilinear} that the bilinear counterpart of this functional is explicitly 
\begin{align}
	\hat{T}_{ab} [ \psi, \psi' ] = \frac{1}{4\pi} \Big\{ \nabla_{(a} \psi \nabla_{b)} \psi' - \tfrac{1}{2} g_{ab} ( \nabla^c \psi \nabla_c \psi' + \mu^2 \psi \psi' ) + \xi \big[ (R_{ab} - \tfrac{1}{2} g_{ab} R) \psi \psi'
	\nonumber
	\\
	~ - (\delta^c_{(a} \delta^d_{b)} - g_{ab} g^{cd}) \nabla_c ( \psi \nabla_d \psi' + \psi' \nabla_d \psi ) \big] \Big\} .
	\label{ThatSc}
\end{align} 
Furthermore, the definition of $\hat{T}_{ab}[\psi,\psi']$ and the conservation of $T_{ab}[\psi]$ imply that
\begin{equation}
	\nabla^b \hat{T}_{ab} [ \psi, \psi ' ] = 0
	\label{tFunctCons}
\end{equation}
for any fields $\psi$ and $\psi'$ which both satisfy \eqref{KleinGordon}. 

Applying \eqref{avDef} and \eqref{Psi2Av} together with these results and the transport equations, the averaged stress-energy tensor associated with $\Psi$ is found to be
\begin{equation}
	\langle T_{ab} \rangle = \frac{ \omega^2 }{ 4\pi }  \left[ \langle \Psi^2 \rangle k_a k_b + \omega^{-1} k_{(a} \Im \left( \calA_0 \nabla_{b)} \bar{\calA}_0 \right) +\mathcal{O}(\omega^{-2}) \right]
	\label{TavZeroth}
\end{equation}
through leading and subleading orders. Neither $\mu$ nor $\xi$ appear explicitly in this expression, although they do appear at the first omitted order. This may be seen by directly computing the average of the trace, 
\begin{align}
	\langle T^{a}{}_{a} \rangle = - \frac{1}{8\pi} \left\{ \mu^2 |\calA_0|^2 + (1- 6 \xi) \left[ |\nabla \calA_0|^2 + \Re (\calA_0 \Box \bar{\calA}_0) \right] \right\} + \mathcal{O}(\omega^{-1}),
	\label{TtrSc}
\end{align}
whose first nonzero terms are suppressed by two powers of $\omega$ with respect to other components of $\langle T_{ab} \rangle$. The latter terms in \eqref{TtrSc} may be seen to vanish for fields in which $\xi =1/6$, which is the value associated with conformal coupling.

The averaged stress-energy tensor reduces to $(\omega^2/8\pi) |\calA_0|^2 k_a k_b$ in the geometric-optics approximation, and this form is almost unchanged at subleading order: Using \eqref{Kscalar} to recall the definition for $\hat{k}_a$, \eqref{TavZeroth} factorizes to
\begin{equation}
	\langle T_{ab} \rangle = \frac{ \omega^2 }{ 4 \pi } \langle \Psi^2 \rangle  \hat{k}_a \hat{k}_b + \mathcal{O}(\omega^{0}) .
	\label{TScalarFact}
\end{equation}
The stress-energy tensor thus retains its geometric-optics form even at subleading order, but with the corrections 
\begin{equation}
|\calA_0|^2 \mapsto |\calA_0 + \omega^{-1} \calA_1|^2, \qquad k_a \mapsto \hat{k}_a.
	\label{scalarMap}
\end{equation}
This confirms that the $\hat{k}_a$ introduced heuristically above does indeed have the interpretation of a corrected propagation direction: Given any observer with 4-velocity $u^a$, the averaged 4-momentum density $\langle p_a \rangle \equiv - \langle T_{ab} \rangle u^b$ seen by that observer is proportional to $\hat{k}_a$. More precisely,
\begin{equation}
	\langle p_a \rangle  = \frac{\omega \omega_o}{4\pi} \langle \Psi^2 \rangle \hat{k}_a + \mathcal{O}(\omega^0),
	\label{pavScalar}
\end{equation}
where
\begin{equation}
	\omega_o \equiv (-u \cdot \hat{k}) \omega
	\label{omegaO}
\end{equation}
is essentially the observed frequency (or its negative if $\hat{k}^a$ is past-directed). This momentum density is always null and its direction does not depend on $u^a$. That direction may however depend on $\omega$, implying that an object's apparent location can depend on the frequency at which it is observed. Moreover, the averaged energy density is given by
\begin{equation}
	\epsilon \equiv \langle T_{ab}\rangle  u^a u^b = \frac{\omega_o^2}{4\pi} \langle \Psi^2 \rangle + \mathcal{O}(\omega^0).
	\label{Iscalar}
\end{equation}
The subleading terms here may be interpreted as due to i) the corrected Doppler effect associated with $\hat{k}_a - k_a$, and ii) the corrected amplitude associated with $\calA_1$.

\subsection{Conservation laws}
\label{Sect:ConsScalar}

A number of conservation laws may be shown to hold for high-frequency scalar fields. One of these has already been noted, namely the conserved current $J_0^a$ defined by \eqref{J0scalar}. This expresses the usual leading-order law of intensity variation expected from geometric arguments involving the focusing or defocusing of optical rays. Given the resemblance of the corrected stress-energy tensor \eqref{TScalarFact} to its geometric-optics counterpart, a similar conservation law might be expected to hold also to subleading order, but with the replacements \eqref{scalarMap}. This is indeed the case. A direct calculation using \eqref{phi0trans} and \eqref{phiNtrans} shows that 
\begin{equation}
	|\calA_0 + \omega^{-1} \calA_1|^2 \hat{k}^a = J_0^a + \omega^{-1} J_1^a + \mathcal{O}(\omega^{-2})
	\label{Jexpand}
\end{equation}
is conserved up to terms of order $\omega^{-2}$. The $\mathcal{O}(\omega^{-1})$ coefficient
\begin{equation}
	J^a_1 \equiv 2\Re(\calA_0 \bar{\calA}_1) k^a - | \calA_0|^2 \nabla^a \mathrm{arg} \, \calA_0
	\label{J1scalar}
\end{equation}
is conserved exactly, a result which is related to intensity variations driven by the different geometrical cross sections associated with the uncorrected and corrected ray congruences determined by $k^a$ and $\hat{k}^a$.

Conservation laws can be associated not only with vector fields, but also with rank-2 symmetric tensor fields. Such laws may be generated systematically 
by noting that $\langle T_{ab} \rangle$ must be conserved at all orders by virtue of \eqref{avDef} and \eqref{tFunctCons}. If the averaged stress-energy tensor is expanded in powers of $\omega^{-1}$ so that
\begin{equation}
	\langle T_{ab} \rangle = \frac{\omega^2}{8\pi} \sum_{n=0}^\infty \omega^{-n} \mathcal{T}_{ab}^n,
	\label{Texpand}
\end{equation}
each coefficient $\mathcal{T}^n_{ab}$ is thus conserved:
\begin{equation}
	\nabla^b \mathcal{T}^n_{ab} = 0.
	\label{TnCons}
\end{equation}
These coefficients propagate without interaction. Comparing \eqref{TavZeroth} and \eqref{Texpand}, the first two examples of conserved tensors generated in this way may be related to the  conserved currents $J^a_0$ and $J^a_1$ via
\begin{align}
	\mathcal{T}_{ab}^0 = |\calA_0|^2 k_a k_b =k_{(a} J^0_{b)},
	\qquad 	
	\mathcal{T}_{ab}^{1}= k_{(a}  \big(  J^1_{b)} - |\calA_0|^2 \nabla_{b)} \arg \calA_0 \big) .
	\label{TcalDef}
\end{align}
These are both trace-free and also transverse in the sense that they vanish when contracted with $k^a$.

Of course, conservation laws like \eqref{TnCons} are most useful in the presence of symmetries, in which case they imply the existence of conserved currents: If $v^a$ is a Killing vector field, 
\begin{equation}
	\mathcal{J}^a_n \equiv \mathcal{T}^n_{ab} v^b
\end{equation}
must be conserved. More than this, for $n=0$ and $n=1$, the vanishing traces of $\mathcal{T}^0_{ab}$ and $\mathcal{T}^1_{ab}$ imply that it is sufficient that $v^a$ be only conformally Killing. In these cases, the currents $\mathcal{J}^a_i$ and $J^a_i$ are related to one another via
\begin{equation}
	\mathcal{J}^a_0 = (k \cdot v) J^a_0, \qquad \mathcal{J}^a_1 = (k \cdot v) J_1^a  - |\calA_0|^2 (\mathcal{L}_v \arg \calA_0) k^a,
\end{equation}
where $\mathcal{L}_v$ denotes the Lie derivative with respect to $v^a$. When appropriate symmetries exist, these currents may be used to compute conserved energies, angular momenta, and so on in finite regions. 

\section{Electromagnetic fields}
\label{Sect:EM}

A WKB ansatz may be used to understand electromagnetic fields just as it can for Klein-Gordon fields. There are at least two interesting ways to proceed: One of these works directly with the field strength $F_{ab}$ \cite{EhlersGeoOptics, Anile, DolanCirc} while the other fixes a gauge and expands a vector potential $A_a$ \cite{Isaacson1, Mashhoon}. The latter approach is adopted here due to its similarity with the Klein-Gordon case. 

\subsection{Geometric optics and its corrections}

Given a vector potential $A_a$, a field strength may be computed using $F_{ab} = 2 \nabla_{[a} A_{b]}$. This automatically solves the Maxwell equation $\nabla_{[a} F_{bc]} = 0$. Restricting to Lorenz gauge, the other Maxwell equation is satisfied in the absence of sources if 
\begin{equation}
	\Box A_a - R_{a}{}^{b} A_b = 0, \qquad \nabla^a A_a = 0.
	\label{MaxwellA}
\end{equation}
While all physical fields here are assumed to be real, it is again convenient to consider a 1-parameter family of complex fields which satisfy these same equations. Letting these have the form
\begin{equation}
	a_a(x;\omega) = e^{i \omega \varphi(x)} \sum_{n=0}^\infty \omega^{-n} \calA_a^n (x),
	\label{Aansatz}
\end{equation}
substitution back into \eqref{MaxwellA} shows that the eikonal equation \eqref{eikonal} does not change: $k_a = - \nabla_a \varphi$ must again be null, at least if $\calA_a^0 \neq 0$ in all regions of interest. This procedure also generates the transport equations
\begin{equation}
	L \calA_a^n = - i (\Box \calA_a^{n-1} - R^{b}{}_{a} \calA_b^{n-1}),
	\label{xPortEM}
\end{equation}
as well as the algebraic constraints
\begin{equation}
	k \cdot \calA_n = -i \nabla \cdot \calA_{n-1} 
	\label{aGauge}
\end{equation}
for all $n \geq 0$, where the transport operator $L$ is given by \eqref{transOp} and we have set $\calA^{-1}_a \equiv 0$ for simplicity. Unlike in the scalar case, the electromagnetic amplitudes are algebraically constrained; not all solutions to the transport equations are physically admissible. Nevertheless, if the constraints \eqref{aGauge} are satisfied on an initial hypersurface, \eqref{xPortEM} guarantees that they remain satisfied along all rays emanating from that hypersurface. 

Solving \eqref{xPortEM} and \eqref{aGauge} for all amplitudes up to some order $m$, the result may be substituted back into \eqref{Aansatz} and the series truncated at that order. This results in an approximation for $A_a = \Re a_a$ which solves both equations in \eqref{MaxwellA} up to terms of order $\omega^{-m}$. However, this does not necessarily imply that the full Maxwell equation $\nabla^b F_{ab} = 0$ is satisfied up to terms of this same order; see \ref{App:Approx}. Despite this, it is straightforward to determine which terms are needed in order to consistently compute different observables up to whichever order is desired.

\subsection{Field strengths}

The vector potential is not particularly interesting on its own. The (possibly complex) field strength $f_{ab} \equiv 2 \nabla_{[a} a_{b]}$ is more directly physical, and differentiating \eqref{Aansatz} shows that this has the form
\begin{equation}
	f_{ab} = - 2i \omega e^{i \omega \varphi} \sum_{n=0}^\infty \omega^{-n} \mathcal{F}_{ab}^n,
	\label{Fansatz}
\end{equation}
where
\begin{equation}
	\mathcal{F}^n_{ab} \equiv k_{[a} \calA^n_{b]} + i \nabla_{[a} \calA_{b]}^{n-1}.
	\label{FnDef}
\end{equation}	
Using \eqref{aGauge}, it follows that the leading-order, geometric-optics field $-2 i \omega e^{i \omega \varphi} k_{[a} \calA_{b]}^0$ is a null 2-form with principal null direction $k^a$. As is standard, the real field strength $F_{ab} = \Re f_{ab}$ can be measured by observing the motions of test charges. At leading order, such accelerations are always orthogonal to the projection of $k^a$ into a particle's rest frame, reflecting the transverse nature of electromagnetic radiation. A ``longitudinal force'' may nevertheless arise at subleading order, as may be seen from
\begin{equation}
	k^a (f_{ab} u^b) = u^a \big[ \nabla^b ( k_a \calA_b^0) - \tfrac{1}{2} (\nabla \cdot k) \calA^0_a \big] e^{i \omega \varphi} + \mathcal{O}(\omega^{-1}),
\end{equation}
where $u^b$ denotes a particle's 4-velocity.

Although we have obtained constraint and transport equations for vector potentials and then derived field strengths from those, it can be useful to note that similar transport equations also hold for the $\mathcal{F}_{ab}^n$ themselves. A straightforward calculation shows that
\begin{equation}
	\mathcal{F}^n_{ab} k^b = - i \nabla^b \mathcal{F}_{ab}^{n-1} , 
	\label{fDotk}
\end{equation}
and
\begin{equation}
	L\mathcal{F}_{ab}^n = - i \left( \Box \mathcal{F}_{ab}^{n-1} + R_{ab}{}^{cd} \mathcal{F}_{cd}^{n-1} + 2 \mathcal{F}_{c[a}^{n-1} R_{b]}{}^{c} \right) .
	\label{Lf}
\end{equation}
Similar equations have also been obtained by Dolan \cite{DolanCirc}. As is more apparent from the treatment there, the combination of curvatures which appear in the transport equations for the $\mathcal{F}_{ab}^n$ are related to the fact that the source-free Maxwell equation $\nabla^b f_{ab} = 0$ may be used to show that
\begin{equation}
	\Box f_{ab} + R_{ab}{}^{cd} f_{cd} + 2f_{c[a} R_{b]}{}^{c} = 0.
\end{equation}
Eqs. \eqref{fDotk} and \eqref{Lf} allow one to understand aspects of the field strength without first computing a vector potential. However, caution must be used in that context as there are solutions to those equations for which $\nabla_{[a} f_{bc]} \neq 0$.

\subsection{Polarization}
\label{Sect:EMpol}

One important difference between scalar and electromagnetic waves is that the latter carry with them a notion of polarization. This can be understood at leading order by factoring $\calA^0_a$ so as to remove any variations which arise even for a scalar field: It is convenient to introduce a ``polarization 1-form'' $e_a$ such that 
\begin{equation}
	\calA_a^0 = \calA_0 e_a,
	\label{AeSplit}
\end{equation}
where $\calA_0$ satisfies the scalar transport equation $L \calA_0 = 0$. It  follows from the $n=0$ case of \eqref{xPortEM} that 
\begin{equation}
	k \cdot \nabla e_a = 0,
	\label{ePar}
\end{equation}
and from the $n=0$ case of \eqref{aGauge} that $k \cdot e= 0$. The polarization is therefore parallel transported along the rays of the field. Moreover, $e \cdot e$ and $e \cdot \bar{e}$ cannot vary along any individual ray. If $e_a$ is not proportional to $k_a$, there is no loss of generality in rescaling $\calA_0$ such that\footnote{If $e_{[a} k_{b]}=0$, it follows from \eqref{FnDef} that $\mathcal{F}_{ab}^0 = 0$. This does not imply, however, that such cases are necessarily unphysical. Nonzero field strengths may be generated at higher orders by zeroth-order amplitudes with this property; see  \ref{Sect:kPol}. We nevertheless assume $e_{[a} k_{b]} \neq 0$ unless otherwise noted.}
\begin{equation}
e \cdot \bar{e} = 1
\label{e2}
\end{equation}
everywhere. Doing so hereafter unless stated otherwise, it follows from \eqref{AeSplit} that the vector expression $g^{ab} \calA_a^0 \bar{\calA}_b^0$ is equal to its scalar analog $\calA_0 \bar{\calA}_0$. We denote both by $|\calA_0|^2$. The leading-order scalar intensity law associated with the conservation of \eqref{J0scalar} thus remains valid also for electromagnetic fields; polarization does not affect intensity in geometric optics.

If $\mathcal{F}^0_{ab}\neq 0$, it is somewhat imprecise to identify $e_a$ as \textit{the} polarization state of the electromagnetic wave, as any modification $e_a \mapsto e_a + \chi k_a$ results in the same leading-order field. It is less ambiguous to say instead that the null 2-form $k_{[a} e_{b]}$ encodes a wave's leading-order polarization state. The space of physical polarization states associated with a nonzero $\mathcal{F}^0_{ab}$ at a point may be identified with the space of 2-forms $k_{[a} e_{b]}$ for which $k \cdot e = 0$ and $e \cdot \bar{e} = 1$, modulo overall phases (which can always be absorbed into redefinitions of $\calA_0$). This space is two-dimensional, so leading-order fields may be characterized by two independent polarization states. Linear polarization may be defined at a point to correspond to cases in which $|e \cdot e|^2 = 1$, which implies that $e_{[a} \bar{e}_b k_{c]} = 0$. Circular polarization may instead be characterized by $e \cdot e = 0$. If a field is linearly or circularly polarized at a point, \eqref{ePar} implies that it retains that characteristic along the entire ray which passes through that point. 

It is convenient for various calculations below to introduce a particular basis of circular polarization states, denoted by $m^a$ and its complex conjugate $\bar{m}^a$. More generally, consider a null tetrad
\begin{equation}
	(k^a, n^a, m^a , \bar{m}^a)
	\label{tetrad}
\end{equation}
which is parallel propagated along the rays tangent to $k^a$, where $m \cdot \bar{m} = - k \cdot n =1$ and all other inner products vanish. There then exist scalars $e_\pm$ and $\chi$ such that
\begin{equation}
	e_a = e_+ m_a + e_- \bar{m}_a + \chi k_a.
	\label{eExpand}
\end{equation}
These parameters remain constant along each ray, and the normalization \eqref{e2} is equivalent to demanding that
\begin{equation}
	|e_+|^2 + |e_-|^2 = 1.
\end{equation}
In terms of these variables, circular and linear polarization states are characterized by 
\begin{equation}
	e_+ e_- =0 \quad \mbox{(circ. pol.)}, \qquad |e_+| = |e_-| =\frac{1}{\sqrt{2}} \quad \mbox{(lin. pol.)}.
	\label{polDefs}
\end{equation}
The term controlled by $\chi$ does not affect $\mathcal{F}^0_{ab}$ and is therefore absent from these expressions.

\subsection{Newman-Penrose scalars and peeling}
\label{Sect:NPEM}

One way to understand the tensorial character of the electromagnetic field beyond leading-order is to introduce a tetrad and to use this to compute the tetrad components of $F_{ab}$. Suppose in particular that a parallel-propagated null tetrad with the form \eqref{tetrad} has been fixed. It is then known that any real 2-form can be completely characterized by the three complex Newman-Penrose scalars \cite{ExactSolns, FrolovRev}
\begin{equation}
	\Phi_0 \equiv F_{ab} k^a m^b, \quad \Phi_1 \equiv \frac{1}{2} F_{ab} (k^a n^b + \bar{m}^a m^b) , \quad \Phi_2 \equiv F_{ab} \bar{m}^a n^b.
	\label{NPEM}
\end{equation}
Inverting these definitions shows that 
\begin{equation}
	F_{ab} = 4 \Re \big[ \Phi_0 \bar{m}_{[a} n_{b]}  + \Phi_1 (n_{[a} k_{b]} + m_{[a} \bar{m}_{b]} )+ \Phi_2 k_{[a} m_{b]} \big].
	\label{FNP}
\end{equation}
The Newman-Penrose scalars thus determine the coefficients of the three terms in this expression. While these scalars have been computed before for high-frequency circularly-polarized fields \cite{DolanCirc}, here we allow for arbitrary polarizations and also state a kind of peeling result which summarizes how the scalars depend on $\omega$.

If the real component of the high-frequency expansion \eqref{Fansatz} for $f_{ab}$ is inserted into the definitions \eqref{NPEM}, the geometric-optics term is found to contribute only to $\Phi_2$. Explicitly,
\begin{equation}
	\Phi_2 = \omega \bar{m}^a \Im \big( \calA^0_a e^{i\omega\varphi} \big) + \mathcal{O}(\omega^0).
	\label{Phi2}
\end{equation}
Using \eqref{AeSplit} and \eqref{eExpand} to expand $e_a$ in terms of $e_\pm$ and $\chi$ shows that this depends on the polarization state via
\begin{equation}
	\Phi_2 = - \frac{1}{2} i \omega |\calA_0| ( e_+ e^{i \omega \hat{\varphi}} - \bar{e}_- e^{-i \omega \hat{\varphi} } ) + \mathcal{O}(\omega^0),
	\label{Phi2pol}
\end{equation}
where $\hat{\varphi}$ is the corrected phase \eqref{phaseCorrect}. It may be interpreted further by noting that its magnitude is
\begin{equation}
	|\Phi_2| = \frac{1}{2} \omega |\calA_0| \big[ 1 - 2 \Re \big( e_+ e_- e^{2i \omega \hat{\varphi}} \big) \big]^{1/2} + \mathcal{O}(\omega^0).
	\label{Phi2Mag}
\end{equation}
Recalling \eqref{polDefs}, the oscillatory term here vanishes only if a wave is circularly polarized. If it is instead linearly polarized, $|\Phi_2|$ oscillates rapidly and passes through zero each time $\omega \hat{\varphi}$ advances by $\pi/2$. 

Moving beyond geometric optics, the constraint and transport equations may be used to show that
\begin{align}
	\Phi_1 = - m^a \bar{m}^b \Re \big( \nabla_a \calA^0_b e^{i \omega \varphi} \big) - \tfrac{1}{2} (\nabla \cdot k) n^a \Re \big( \calA_a^0 e^{i \omega \varphi} \big) + \mathcal{O}(\omega^{-1}).
	\label{Phi1}
\end{align}
The second term here projects out any dependency on the coefficient $\chi$ in \eqref{eExpand}; while such terms are not necessarily unphysical, they first contribute in this context via a subleading correction to $\Phi_2$. Expanding $\Phi_1$ for a general polarization state,
\begin{align}
	\Phi_1 = - \frac{1}{2} m^a |\calA_0| \big\{ e_+ \big[ \bar{m}^b \nabla_a m_b + \nabla_a \ln (e_+ \calA_0) \big] e^{i \omega \hat{\varphi}} +\bar{e}_- \big[ \bar{m}^b \nabla_a m_b
	\nonumber
	\\
	~ +  \nabla_a \ln (\bar{e}_- \bar{\calA}_0 ) \big] e^{-i \omega \hat{\varphi} }  \big\} + \mathcal{O}(\omega^{-1}).
	\label{Phi1pol}
\end{align}
Again, rapid oscillations in the magnitude of this quantity disappear for circularly-polarized waves: Temporarily specializing to that case and choosing $m^a$ such that $e_+ = 1$ and $e_- = 0$,
\begin{align}
	\Phi_1 = - \frac{1}{2} m^a |\calA_0| \big( \bar{m}^b \nabla_a m_b + \nabla_a \ln \calA_0 \big) e^{i \omega \hat{\varphi}} + \mathcal{O}(\omega^{-1})	.
\end{align}
Overall, $\Phi_1$ may be viewed as measuring the degree by which the field varies in directions transverse to its leading-order direction of propagation.

The last of the Newman-Penrose scalars can arise at the same order as $\Phi_1$, and is given by
\begin{align}
	\Phi_0 = - \sigma \bar{m}^a \Re \big( \calA^0_a e^{i \omega \varphi} \big) + \mathcal{O}(\omega^{-1}),
	\label{Phi0}
\end{align}
where 
\begin{equation}
	\sigma \equiv - m^a m^b \nabla_a k_b
	\label{shearDef}
\end{equation}
denotes the complex shear of the ray congruence tangent to $k^a$. Despite its definition, the magnitude of $\sigma$ depends only on $k^a$ and not $m^a$:
\begin{equation}
	|\sigma|^2 = \frac{1}{2} \big[ \nabla_a k_b \nabla^a k^b - \frac{1}{2} (\nabla \cdot k)^2 \big].
\end{equation}
Expanding \eqref{Phi0} for a general polarization state shows that
\begin{equation}
	\Phi_0 = - \frac{1}{2} \sigma |\calA_0| ( e_+ e^{i \omega \hat{\varphi}} + \bar{e}_- e^{-i \omega \hat{\varphi}} ) + \mathcal{O}(\omega^{-1}),
	\label{Phi0pol}
\end{equation}
the magnitude of which is
\begin{equation}
	|\Phi_0| = \frac{1}{2} |\sigma| |\calA_0| \big[ 1 + 2 \Re \big( e_+ e_- e^{2i \omega \hat{\varphi}} \big) \big]^{1/2} + \mathcal{O}(\omega^{-1}). 
\end{equation}
Here too, rapid variations disappear for circularly-polarized waves. The information encoded in the given expression for $\Phi_0$ is not significantly different from that given by $\Phi_2$ at leading order, except that $\Phi_0$ scales differently with $\omega$ and is multiplied by $\sigma$.

There is a sense in which $\sigma \neq 0$ is generic; the Goldberg-Sachs theorem \cite{ExactSolns, FrolovRev} states that at least in Ricci-flat spacetimes, shear-free null geodesic congruences do not exist unless the metric is algebraically special. Even in spacetimes where shear-free rays may exist, they are special. Summarizing \eqref{Phi2}, \eqref{Phi1}, and \eqref{Phi0}, the generic frequency scalings associated with the electromagnetic Newman-Penrose scalars are
\begin{equation}
	\begin{gathered}
		\Phi_i = \mathcal{O}(\omega^0), \qquad i = 0,1,
		\\
		\Phi_2 = \mathcal{O}(\omega).
	\end{gathered}
	\label{Peeling0}
\end{equation}
This is a kind of local peeling result which describes the relative significance of the terms in \eqref{FNP}.

A somewhat simpler scaling arises if $\sigma$ vanishes, or is at least negligible. While these cases are not generic in the sense described above, they include a number of important examples. For example, the rays associated with a radiating point particle are shear-free in any conformally-flat spacetime. In more general geometries which are at least asymptotically flat, the (nonzero) shear associated with a radiating particle would decay rapidly with distance. Regardless, setting $\sigma = 0$ in \eqref{Phi0} shows that \eqref{Peeling0} simplifies to
\begin{equation}
	\Phi_i = \mathcal{O}(\omega^{i-1}), \qquad i = 0,1,2
	\label{Peeling}
\end{equation}
in a shear-free context. The three terms in \eqref{FNP} thus fall off at different rates as $\omega \to \infty$. Other peeling results in the literature \cite{PenrosePeeling,HoganEllisCosmology} obtain superficially-similar scalings, except in inverse powers of distance instead of frequency. However, those statements depend on a specific choice of boundary conditions. Eq. \eqref{Peeling} does not. Still, the two results are not unrelated: Noting that a field radiated by a compact source in an asymptotically-flat spacetime encounters less curvature, less ray expansion, and less shear as it propagates outwards, all lengthscales tend to infinity at large distances. Moreover, ratios of successive terms in a high-frequency expansion may be estimated using powers of $(\omega \ell)^{-1}$, with $\ell$ an appropriate lengthscale. Combining these statements implies that any $\omega$ is ``large'' at sufficiently large distances. High-frequency expansions may thus be used to derive large-distance expansions in this context.

\subsection{Directions associated with the field}
\label{Sect:EMPND}

One of the most basic characteristics of the geometric-optics field is its propagation direction $k^a$, and it is natural to ask how this might be corrected at finite frequencies. In the scalar context, the factorization \eqref{phaseGuess} of the leading-order field suggested the corrected direction $\hat{k}^a$, as given by \eqref{Kscalar}, and the physical interpretation of this guess was confirmed\footnote{Other criteria may nevertheless be used to be obtain other generalizations of $k^a$. For example, \eqref{scalarForce} suggests a different (though rapidly varying) direction based on the forces which act to test charges.} by the factorization \eqref{TScalarFact} of the field's stress-energy tensor, and especially by the momentum density \eqref{pavScalar}. Unfortunately, the same simple arguments fail in the electromagnetic context. The problem is essentially that an electromagnetic field has several scalar components, and each of these may suggest a different effective phase. Worse, it is shown below that the electromagnetic stress-energy tensor does not remain in geometric-optics form beyond leading order: While the direction of the subleading 4-momentum density is indeed corrected relative to $k^a$, that correction can be observer-dependent for an electromagnetic field. It thus appears that although geometric optics remains ``essentially valid'' even at subleading order for stress-energy tensors associated with Klein-Gordon fields, the same cannot be said for electromagnetic fields. 

Despite this, a considerable literature has grown up around ascribing helicity-dependent corrections to propagation directions associated with circularly-polarized fields in curved spacetimes \cite{ShoomSpinOptics, YooSpin, Duval2017, Bailyn1977, Bailyn1981, Gosselin2007, DuvalFermat}. In some of these cases \cite{ShoomSpinOptics, YooSpin}, different components of the electromagnetic field are evaluated with respect to a certain frame and then factorized to motivate corrections to the eikonal equation. It is not made clear how these results are directly interpretable as propagation directions, and in any case they depend upon the chosen frame. Other approaches note that there are cases in which the Mathisson-Papapetrou equations govern the linear and angular momenta of a ``photon,'' and that its trajectory may be deduced by combining these equations with an appropriate centroid (or spin supplementary) condition. While the momenta $P^a$ and $S^{ab}$ of suitable classical wavepackets are indeed governed by the Mathisson-Papapetrou equations, imposing a supplementary condition such as $P_a S^{ab} = 0$ (as in, e.g., \cite{Duval2017}) may be shown to fail even for plane-fronted waves in flat spacetime; that condition constrains only one component of the centroid, not three\footnote{This follows from applying the standard definitions for $P^a$ and $S^{ab}$ (see, e.g., \cite{MTW}) to a stress-energy tensor proportional to $k^a k^b$, where $k^a$ is null and constant. Separately, it may be seen directly that the equations of motion in \cite{Duval2017} are ill-defined in flat spacetime. This is explained there by saying that massless spinning particles are ``delocalized'' in that case. However, narrow beams in flat spacetime clearly \textit{are} localizable; the connection with classical wavepackets is therefore unclear.}. Different spin-supplementary conditions are motivated in \cite{Bailyn1977,Bailyn1981} and shown to imply that spinning massless particles move on null geodesics. 

The many different approaches and conclusions in these papers and others appear to be symptoms of the fact that it is not necessarily meaningful to define \textit{a} direction of propagation beyond leading order. While momentum densities and beam centroids do shift at finite frequencies, it can be misleading to ascribe these and other phenomena to a single ``corrected propagation direction;'' different directions might arise for different phenomena. The point of view adopted here is that the single propagation direction associated with geometric optics splits into two at finite frequencies. Both directions must be taken into account in order to describe observables beyond geometric optics\footnote{Geometric intuition must still be treated with caution. Even with two directions at hand, most results cannot be described as an incoherent sum of two geometric-optics expressions with different propagation directions. While the directions we consider are well-defined, it is debatable whether or not it is useful to refer to them as \textit{propagation} directions.}.

The directions we consider first are the real principal null directions of $F_{ab}$. These are known to be locally determined, to possess clear physical interpretations, and to be well-defined for any nonzero field, even in the absence of any approximation \cite{FrolovRev, PenrosePeeling, SyngeSR, Hall}. The principal null directions are essentially the null eigenvectors of $F_{ab}$. More precisely, a principal null \textit{direction} is defined to be a congruence tangent to any nonzero null vector field $k'^a$ which satisfies
\begin{equation}
	k'^a F_{a}{}^{[b} k'^{c]} = 0 . 
	\label{PNDdef}
\end{equation} 
We refer to any such vector field as a principal null \textit{vector}. Multiplying one principal null vector by any nonzero scalar results in another principal null vector but the same principal null direction. Besides their direct interpretation as eigenvectors of $F_{ab}$, principal null vectors are also eigenvectors of a field's (full, non-averaged) stress-energy tensor. As mentioned above, the geometric-optics field strength admits exactly one principal null direction, namely that determined by the ray congruence tangent to $k^a$. At higher orders, this single direction generically splits into two.

Finite-frequency corrections to the principal null directions may be found using the Newman-Penrose scalars discussed in Sect. \ref{Sect:NPEM}. First consider a null tetrad 
\begin{equation}
	(k'^a,n'^a,m'^a,\bar{m}'^a)
	\label{tetradPrime}
\end{equation}	
which is normalized in the same way as the unprimed tetrad \eqref{tetrad}. Adapting a statement regarding the principal null directions of Weyl tensors in \cite{Chandra}, it may be shown that $\Phi'_0 \equiv F_{ab} k'^a m'^b = 0$ if and only if $k'^a$ is a principal null vector. All real principal null directions may therefore be found by finding those tetrads  \eqref{tetradPrime} for which $\Phi'_0 = 0$. If the unprimed tetrad \eqref{tetrad} is taken as an initial guess, a rotation may be applied in order to generate a new, primed tetrad with this property. Appropriate rotations may be parametrized by a complex scalar $z$, whence
\begin{equation}
	k'^a = k^a + |z|^2 n^a + \bar{z} m^a + z \bar{m}^a, \qquad n'^a = n^a, \qquad m'^a = m^a + z n^a.
	\label{tetradRotate}
\end{equation}
All inner products between the tetrad components are preserved by these transformations. Furthermore,
\begin{equation}
	\Phi'_0 = \Phi_0 + 2 z \Phi_1 + z^2 \Phi_2 = 0
	\label{zPND}
\end{equation}
if $k'^a$ is to satisfy \eqref{PNDdef}. This is a quadratic equation for $z$, with solutions
\begin{equation}
	z = \frac{1}{\Phi_2} \left[ - \Phi_1 \pm \left( \Phi_1^2 - \Phi_0 \Phi_2 \right)^{1/2} \right],
	\label{zSoln}
\end{equation}
at least if $\Phi_2 \neq 0$. Each $z$ determines, via \eqref{tetradRotate}, a real principal null direction associated with $F_{ab}$. 

The strategy now is to solve \eqref{zPND} using the $\Phi_i$ computed in Sect. \ref{Sect:NPEM}. However, as mentioned there, $\Phi_2$ rapidly oscillates through zero for linearly-polarized waves. Stated differently, the field vanishes periodically and the eigenvector problem is ill-defined wherever it does so. These difficulties may be avoided for circularly polarized waves, and it is only in that case for which we explicitly evaluate the $k'_a$. Suppose in particular that the $m^a$ component of the unprimed tetrad is chosen to coincide with $e_a$, so $e_+=1$ and $e_-=0$. It then follows from \eqref{Phi2pol}, \eqref{Phi0pol}, and \eqref{zSoln} that
\begin{align}
	z = \pm ( i \sigma/ \omega)^{1/2} + \mathcal{O}(\omega^{-1}).
	\label{zExpand}
\end{align}
Substituting this back into \eqref{tetradRotate} shows that the principal null vectors for a wave with circular polarization $e_a = m_a$ are given by
\begin{equation}
	k'^a = k^a \pm 2 \Re \big[ \left(-i \bar{\sigma} / \omega \right)^{1/2} m^a \big] + \mathcal{O}(\omega^{-1}).
	\label{pertPND}
\end{equation}
Unlike all other quantities considered in this paper, the first correction here scales like $\omega^{-1/2}$ instead of an integer power of $\omega^{-1}$. In this sense, principal null directions are particularly sensitive to finite-wavelength effects. Similar dependencies on square roots of expansion parameters have been noted before for the principal null directions associated with perturbed Weyl tensors in Petrov type-D backgrounds \cite{PertPND1,PertPND2}.

It is clear from \eqref{pertPND} that the single leading-order principal null direction splits into two whenever $\sigma \neq 0$. This dependence on the shear is reminiscent of---although different from---Robinson's theorem \cite{Robinson, FrolovRev}, which non-perturbatively relates shear-free null geodesic congruences to null electromagnetic fields (i.e., fields which admit only one principal null direction).  This theorem implies in particular that if $\sigma \neq 0$, there does not exist an exact Maxwell field whose principal null congruence is tangent to $k^a$. One might therefore suspect that the 1-parameter family of fields associated with the high-frequency approximation cannot all be null if the leading-order approximation for their principal null vectors has nonzero shear. However, it does not appear to imply a particular order at which nonzero shear forces the principal null directions to split.

If $\sigma = 0$, the Newman-Penrose scalars satisfy the peeling result \eqref{Peeling} and the first correction to the principal null directions may be seen from \eqref{zSoln} to scale like $\omega^{-1}$, not $\omega^{-1/2}$. Computing this correction explicitly would require evaluating $\Phi_0$ to one higher order than in \eqref{Phi0}, which we do not do. Nevertheless, the principal null directions may be seen to again split into two, except in special cases where $\Phi_1^2 = \Phi_0 \Phi_2$. Indeed, this latter condition is sufficient (at all orders) to imply that there exists only a single principal null direction.

Although we have explicitly computed principal null directions only through $\mathcal{O}(\omega^{-1/2})$ and only for circularly-polarized fields, closely-related directions are determined below, through $\mathcal{O}(\omega^{-1})$ and for general polarization states; cf. \eqref{kDefem}. These are the eigenvectors of the field's averaged stress-energy tensor. The distinction between these directions and the principal null directions may be seen by noting that if a real null vector field is an eigenvector of $f_{ab}$, it is also an eigenvector of $F_{ab}$, $T_{ab}[F_{cd}]$, and $\langle T_{ab} \rangle$. However, while real eigenvectors of $F_{ab}$ are also eigenvectors of $T_{ab} [ F_{cd}]$, they are not necessarily eigenvectors of $f_{ab}$ or $\langle T_{ab} \rangle$. Despite this difference in general, the eigenvectors of $\langle T_{ab} \rangle$ calculated below do agree with the principal null vectors for circularly-polarized fields through $\mathcal{O}(\omega^{-1/2})$.

\subsection{Stress-energy tensors and other quadratic observables}
\label{Sect:EMT}

As in the scalar case considered in Sect. \ref{Sect:ScalarObs} above, a high-frequency electromagnetic field may be characterized via averages of various quantities which are quadratic in that field. The simplest such quantity is simply the squared-magnitude of the vector potential,
\begin{equation}
	\langle A^2 \rangle \equiv \frac{1}{2} a^b \bar{a}_b =\frac{1}{2} |\calA_0 + \omega^{-1} \calA_1|^2 + \mathcal{O}(\omega^{-2}).
	\label{A2av}
\end{equation}
The average here is defined by the first equality and follows the prescription given by \eqref{Sbilinear} and \eqref{avDef}. The result is not essentially different from its scalar counterpart \eqref{Psi2Av}.

More interesting are the counterparts of $\langle |\nabla\Psi|^2\rangle$. This single average in the scalar context generalizes to two averages for electromagnetic fields, namely those of $F_{ab} F^{ab}$ and $F_{ab} {}^* F^{ab} =\frac{1}{2} \epsilon_{abcd} F^{ab} F^{cd}$. Both of these quantities vanish for null fields, and therefore vanish in geometric optics. Their averages are in fact suppressed by two powers of $\omega$ relative to generic components of $\langle F_{ab} F_{cd} \rangle = \mathcal{O}(\omega^2)$: Using \eqref{xPortEM}, \eqref{aGauge}, and \eqref{Fansatz},
\begin{align}
	\langle F_{ab} F^{ab} \rangle = \Re \big[ \calA^a_0 \big( \Box \bar{\calA}^0_a - 2 \nabla_a \nabla_b \bar{\calA}^b_0 - R_{ab} \bar{\calA}^b_0 \big) \big] - |\nabla \cdot \calA_0|^2 
	\nonumber
	\\
	~ +2 \nabla^a \calA^b_0 \nabla_{[a} \bar{\calA}_{b]}^0 + \mathcal{O}(\omega^{-1}) , 
\end{align}
and
\begin{align}
	\langle F_{ab} {}^{*}F^{ab} \rangle = \epsilon_{abcd} \big[ 2 k^a \Im (\calA^b_1 \nabla^c \bar{\calA}^d_0 - \bar{\calA}_0^b \nabla^c \calA^d_1) + \nabla^a \calA_0^b \nabla^c \bar{\calA}_0^d \big] + \mathcal{O}(\omega^{-1}).
\end{align}
Note that the non-averaged versions of these quantities can be significantly larger when $\sigma \neq 0$; they are generically of order $\omega^1$ rather than $\omega^0$ \cite{DolanCirc}.
%\begin{align}
%	\langle F_{ab} {}^{*}F^{ab} \rangle 
%	=  2 \omega \epsilon_{abcd} k^a  \Im ( \calA_0^b \nabla^c \bar{\calA}^d_0) + \mathcal{O}(\omega^{0} ).
%\end{align}

Other observables associated with an electromagnetic field can be constructed from its stress-energy tensor
\begin{equation}
	T_{ab} [f_{cd} ] = \frac{1}{4\pi} \left( f_{ac} f_{b}{}^{c} - \frac{1}{4} g_{ab} f_{cd} f^{cd} \right) .
\end{equation}
As in the scalar case, it is convenient to use \eqref{Sbilinear} to obtain from this the bilinear functional
\begin{equation}
	\hat{T}_{ab}[f_{cd},f'_{ef}] \equiv \frac{1}{4\pi} \left( f_{(a|c|} f'_{b)}{}^{c} - \frac{1}{4} g_{ab} f^{cd} f'_{cd} \right) ,
\end{equation}
which is conserved when $f_{ab}$ and $f'_{ab}$ both satisfy Maxwell's equations. If \eqref{avDef} is used to define an averaged stress-energy tensor, the above expansions together with \eqref{A2av} and the constraint and transport equations imply that
\begin{align}
	\langle T_{ab} \rangle = \frac{ \omega^2 }{ 4\pi } \Big\{ \langle A^2 \rangle k_a k_b + \omega^{-1} \Im \Big[ \calA^c_0 \nabla_{(a} \big( k_{b)} \bar{\calA}^0_c \big)  
	 - \nabla_c \big( \calA_0^c \bar{\calA}_{(a}^0 k^{}_{b)}  \big) \Big] 
	 \nonumber
	 \\
	  ~+ \mathcal{O}(\omega^{-2}) \Big\}.
	\label{TemExpand}
\end{align}
This reduces to $(\omega^2/8\pi) |\calA_0|^2 k_a k_b$ at leading order, which is identical to the leading-order term in the scalar-field stress-energy \eqref{TavZeroth}. While a similar form is retained for scalar fields even at subleading order, this is not necessarily the case in electromagnetism; polarization effects generically conspire to make the subleading electromagnetic stress-energy genuinely different from its leading-order counterpart.

This may be seen by factorizing $\langle T_{ab} \rangle$. One approach has been discussed by Dolan \cite{DolanCirc} in the circularly-polarized case, who found that if $m_a$ is chosen to coincide with the polarization direction, the averaged stress-energy tensor can be written, through subleading order, as something proportional to $K_a K_b$, where $K_a = k_a + \mathcal{O}(\omega^{-1})$ is null, plus a correction proportional to $\omega^{-1} \Im (i \bar{\sigma} m_a m_b)$. While the $K_a K_b$ term has a clear interpretation in this representation, the remainder does not. A different approach is adopted here. Allowing for general polarization states, $\langle T_{ab} \rangle$ may be written in terms of \textit{two} null vectors---its eigenvectors. Inspired by the principal null vectors \eqref{pertPND}, one might expect that these eigenvectors differ from one another by a $\sigma$-dependent term which scales like $\omega^{-1/2}$. This is indeed the case: Eq. \eqref{TemExpand} may be rewritten as
\begin{equation}
	\langle T_{ab} \rangle = \frac{\omega^2}{4\pi} \langle A^2 \rangle \left[ \hat{k}^+_{(a} \hat{k}^-_{b)} - \tfrac{1}{4}  g_{ab} ( \hat{k}_+ \cdot \hat{k}_-) + \mathcal{O}(\omega^{-2}) \right] ,
	\label{TemFact}
\end{equation}
where
\begin{align}
	\hat{k}_a^\pm \equiv \hat{k}_a \pm 2 \Re ( \bar{z} m_a ) + |z|^2 n_a - \omega^{-1} \Im \big[ \bar{e}_a e \cdot \nabla \ln |\calA_0|^2 + \nabla^b (\bar{e}_a e_b) 
	\nonumber
	\\
	~ - e^b \nabla_a \bar{e}_b - (2 g_{ab} + k_{a} n_{b}) n_c e^d \bar{e}^{(b} \nabla^{c)} k_d \big],
	\label{kDefem}
\end{align}
$\hat{k}_a$ is the scalar propagation direction \eqref{Kscalar}, and
\begin{equation}
	z = \big[ ( |e_+|^2 - |e_-|^2 ) i \sigma /\omega \big]^{1/2}.
\label{zEM}
\end{equation}
The $\hat{k}^\pm_a$ agree with the principal null vectors \eqref{pertPND} at least in the circularly-polarized, $\mathcal{O}(\omega^{-1/2})$ context in which the latter were computed. In general, the eigenvectors here are null through the relevant order:
\begin{equation}
	\hat{k}_\pm \cdot \hat{k}_\pm = \mathcal{O}(\omega^{-2}). 
\end{equation}
Moreover, $\hat{k}_+ \cdot \hat{k}_- = - 4 |z|^2 + \mathcal{O}(\omega^{-2})$. 

If $\sigma = 0$ or if a wave is linearly polarized, it follows from \eqref{polDefs}, \eqref{kDefem}, and \eqref{zEM} that $z=0$ and both eigenvectors coincide. In these cases, $\langle T_{ab} \rangle$ simplifies to a geometric-optics form with only one relevant null vector:
\begin{equation}
	\langle T_{ab} \rangle = \frac{\omega^2}{4\pi} \langle A^2 \rangle  \big[ \hat{k}^+_{a} \hat{k}^+_{b} + \mathcal{O}(\omega^{-2}) \big].
	\label{TemSimp}
\end{equation}
This simplifies even further in the linearly-polarized case where $\chi = 0$ in \eqref{eExpand}, in which case $\hat{k}^+_a = \hat{k}^-_a = \hat{k}_a$; the single effective electromagnetic propagation direction reduces to its scalar counterpart $\hat{k}_a$. At least at the level of averaged stress-energy tensors, linearly-polarized electromagnetic fields for which $\chi = 0$ thus behave very similarly to scalar fields, even at one order beyond geometric optics\footnote{Some differences remain in the sense that the $|\calA_0 + \omega^{-1} \calA_1|^2$ which appear in $\langle A^2 \rangle$ and $\langle \Psi^2 \rangle$ can behave somewhat differently for scalar versus vector amplitudes. This is discussed  in Sect. \ref{Sect:HighRel} below. Furthermore, if an electromagnetic field is linearly polarized but $\chi \neq 0$, the only change to these statements is that the component of the propagation direction proportional to $k_a$ might change: $\hat{k}^+_a = \hat{k}^-_a = \hat{k}_a + \omega^{-1} (\ldots) k_a$.}. The additional complication of the generic electromagnetic problem may therefore be dropped in these cases.

Again allowing for arbitrary polarization states, suppose that there is a timelike observer with 4-velocity $u^a$. Given that there may be two relevant propagation directions, it is convenient to define \textit{two} measurable frequencies by analogy with \eqref{omegaO}, namely
\begin{equation}
	\omega_o^\pm \equiv (- u \cdot \hat{k}^\pm) \omega.
\end{equation}
In terms of these quantities, it follows from \eqref{TemFact} that the averaged momentum density seen by the observer is
\begin{align}
	\langle p_a \rangle =  \frac{ \omega_o^+ \omega_o^- }{8\pi} \langle A^2 \rangle  \Big[ ( \omega / \omega_o^+ ) k^+_a + ( \omega /\omega^-_o ) k^-_a  
	\nonumber
	\\
	~ + \tfrac{1}{2} (\omega^2/ \omega_o^+ \omega_o^- )  (\hat{k}^+ \cdot \hat{k}^-) u_a \Big] + \mathcal{O}(\omega^0) .
	\label{pavEM}
\end{align}
This is a linear combination of momenta in the two propagation directions $\hat{k}^\pm_a$, together with an ``interference term'' proportional to the observer's 4-velocity. Unlike in geometric optics, the direction of $\langle p_a \rangle$ may depend on $u^a$ at this order. However, it remains null in the sense that
\begin{equation}
	\langle p_a \rangle \langle p^a \rangle = \mathcal{O}(\omega^2),
\end{equation}
which is to be compared with $\langle p_a \rangle \langle p_b \rangle = \mathcal{O}(\omega^4)$. Eq. \eqref{pavEM} may be contracted with $-u^a$ to also yield the observed energy density
\begin{equation}
	\epsilon = \frac{\omega_o^+ \omega^-_o }{4\pi} \langle A^2 \rangle \left[ 1 -  \left| |e_+|^2 - |e_-|^2 \right| \left( \frac{ \omega |\sigma| }{  \omega_o^+ \omega_o^- } \right)  \right] + \mathcal{O}(\omega^0).
\end{equation}
This differs from its scalar counterpart \eqref{Iscalar} in two significant ways. First, the scalar prefactor $\omega_o^2$ is replaced by $\omega_o^+ \omega^-_o$ in the electromagnetic energy density, which is the square of the geometric average of the two effective frequencies associated with the electromagnetic wave. Second, the overall expression is reduced in magnitude by a term which depends on the dimensionless ratio $\omega |\sigma| / \omega_o^+ \omega_o^-$. As stated more generally above, both of these distinctions disappear if the field is linearly polarized or if $\sigma = 0$.

\subsection{Conservation laws} 

As in the scalar case, various conservation laws may be associated with high-frequency electromagnetic fields. Most directly, the conserved current $J_0^a$ defined by \eqref{J0scalar} is preserved as-is. An appropriate analog of $J_1^a$ differs from its scalar counterpart \eqref{J1scalar} mainly by the addition of a polarization-dependent term; for electromagnetic fields,
\begin{equation}
	J_1^a \equiv 2 \Re ( \calA^0 \cdot \bar{\calA}^1 ) k^a - |\calA_0|^2 \big( \nabla^a \arg \calA_0 + i e^b \nabla^a \bar{e}_b \big).
	\label{J1em}
\end{equation}
This is real. The conservation of $J_0^a$ and $J_1^a$ implies that
\begin{equation}
	J_0^a + \omega^{-1} J_1^a = |\calA_0 + \omega^{-1} \calA_1|^2 ( \hat{k}^a - i \omega^{-1} e^b \nabla^a \bar{e}_b ) + \mathcal{O}(\omega^{-2})
\end{equation}
is conserved as well, which is the electromagnetic analog of \eqref{Jexpand}. Physically, it may be interpreted as a correction to the leading-order area-intensity law. Note however that the effective propagation direction $\hat{k}^a - i \omega^{-1} e^b \nabla^a \bar{e}_b$ which appears here is different in general from the $\hat{k}^\pm_a$ which arise in the averaged stress-energy tensor.

By the same arguments as in Sect. \ref{Sect:ConsScalar}, an infinite number of separately-conserved, rank-2 symmetric tensors may be generated by expanding $\langle T_{ab} \rangle$ as in \eqref{Texpand}; each coefficient $\mathcal{T}^n_{ab}$ which appears in that expansion is conserved. By comparison with \eqref{TemExpand}, the scalar-field $\mathcal{T}^{ab}_0$ defined by \eqref{TcalDef}  is unchanged for electromagnetic fields. However, $\mathcal{T}^{ab}_1$ is replaced by
\begin{align}
	\mathcal{T}^{ab}_1 \equiv 2 \Re ( \calA^0 \cdot \bar{\calA}^1 ) k^{a} k^b + 2 \Im \Big[ \calA^c_0 \nabla^{(a} \big( k^{b)} \bar{\calA}^0_c \big)  
	 - \nabla_c \big( \calA_0^c \bar{\calA}^{(a}_0 k^{b)}  \big) \Big].
	\label{T1EM}
\end{align}
This is transverse in the sense that it vanishes when contracted with $k_b$. Noting that all of the $\mathcal{T}^n_{ab}$ are trace-free, any conformal Killing field $v^a$ which might exist generates the infinite number of conserved currents $\mathcal{T}_n^{ab} v_b$.

Exact Maxwell fields are known to also admit a large number of conservation laws which are not of the types discussed here \cite{MaxwellCons, SenovillaCons, Andersson2017}. While there is no obstacle to also expanding these at high frequencies, their physical interpretations are less clear.

\section{Gravitational waves}
\label{Sect:Grav}

Our final application for the high-frequency approximation is concerned with weak gravitational waves in general relativity. These are taken to be linear perturbations on a background spacetime whose metric $g_{ab}$ satisfies the vacuum Einstein equation 
\begin{equation}
	R_{ab} = \Lambda g_{ab} ,
	\label{Einstein}
\end{equation}
perhaps in the presence of a cosmological constant $\Lambda$. While it can be interesting to consider non-vacuum backgrounds as well, the matter which is necessarily present in those cases would be perturbed by passing gravitational waves, and the details of those perturbations would depend on the precise nature of the matter involved \cite{Szekeres,EhlersDust,EhlersGenWKB}.

\subsection{Geometric optics and its corrections}

The geometric-optics limit and its corrections may be derived for gravitational waves in almost the same way as for scalar or electromagnetic waves. We begin by imposing the Lorenz gauge condition 
\begin{equation}
	\nabla^b ( H_{ab} - \tfrac{1}{2} g_{ab} H^{c}{}_{c} ) = 0
	\label{LorenzGrav}
\end{equation}
on the real metric perturbation $H_{ab} = H_{(ab)}$, and with this fixed, the linearized Einstein equation reduces to
\begin{equation}
	\Box H_{ab} + 2 R_{a}{}^{c}{}_{b}{}^{d} H_{cd} = 0.
	\label{linEins}
\end{equation}
The high-frequency ansatz then consists of the introduction of a complex metric perturbation $h_{ab}$ which admits the asymptotic expansion
\begin{equation}
	h_{ab} (x;\omega) = e^{i \omega \varphi(x)} \sum_{n=0}^\infty \omega^{-n} \calA^n_{ab}(x).
	\label{hAnsatz}
\end{equation}
Demanding that $h_{ab}$ satisfy \eqref{LorenzGrav} and \eqref{linEins}, the first result which is obtained when substituting \eqref{hAnsatz} into these equations is that the eikonal equation \eqref{eikonal} is unchanged; $k_a = - \nabla_a \varphi$ must be null with respect to the background metric. Defining $\calA_{ab}^{-1} = \calA_{ab}^{-2} \equiv 0$ in order to simplify the notation here and below, the gauge condition is seen to impose the algebraic constraints 
\begin{equation}
	( \delta^c_a \delta^d_b - \tfrac{1}{2} g_{ab} g^{cd} ) ( k^b \calA^n_{cd} + i \nabla^b \calA^{n-1}_{cd} ) = 0.
	\label{hGauge}
\end{equation}
Similarly, the gauge-fixed Einstein equation implies the transport equations
\begin{equation}
	L \calA_{ab}^n = - i ( \Box \calA_{ab}^{n-1} + 2 R_{a}{}^{c}{}_{b}{}^{d} \calA_{cd}^{n-1}),
	\label{hTransport}
\end{equation}
where $L$ is the transport operator \eqref{transOp}. These results hold for all $n \geq 0$. High-frequency metric perturbations $H_{ab} = \Re h_{ab}$ may be constructed by solving \eqref{hGauge} and \eqref{hTransport} for the amplitudes $\calA^n_{ab}$ and then substituting the results into \eqref{hAnsatz}.

\subsection{Curvature perturbations}

While it is possible to analyze gravitational waves directly at the level of the metric perturbation $H_{ab}$, it is often useful to instead consider first-order perturbations $\delta R_{abcd}$ of the Riemann tensor. This is more closely connected to many observables and is also less sensitive to gauge ambiguities. Unlike the electromagnetic field strength, curvature perturbations can depend on the choice of gauge: Any vector field $v^a$ may be used to generate a first-order gauge transformation in which
\begin{equation}
	H_{ab} \mapsto H_{ab} + \mathcal{L}_v g_{ab}, \qquad \delta R_{abcd} \mapsto \delta R_{abcd} + \mathcal{L}_v R_{abcd}.
\end{equation}
It is clear that the latter expression here is independent of $v^a$ only if the background is flat. However, it can make sense to restrict to gauge vectors which do not depend on $\omega$, or at least those for which $v^a$ and $\nabla_a v^b$ remain bounded as $\omega \to \infty$. In these cases, there is a sense in which curvature perturbations are gauge-invariant at leading and subleading orders \cite{Isaacson1}.

Using doubled square brackets to denote independent antisymmetrizations over the outer and inner pairs of indices [so, e.g., $f_{[a[bc]d]} = \frac{1}{2} ( f_{a[bc]d} - f_{d[bc]a})$ for any $f_{abcd}$], the linearized perturbation $\delta R_{abcd}$ may be computed as the real component of
\begin{equation}
	\delta r_{abcd} = - 2 \nabla_{[a} \nabla_{[c} h_{d]b]} + R_{ab [c}{}^{f} h_{d]f} .
	\label{Rdef}
\end{equation}
It is convenient to expand this in powers of $\omega^{-1}$, defining coefficients $\mathcal{R}^n_{abcd}$ such that
\begin{equation}
	\delta r_{abcd} = - 2 \omega^2 e^{i \omega \varphi} \sum_{n=0}^\infty \omega^{-n} \mathcal{R}^n_{abcd}.
	\label{rAnsatz}
\end{equation}
Eqs. \eqref{hAnsatz} and \eqref{Rdef} show that for all $n \geq 0$, these coefficients are related to the metric-perturbation amplitudes $\calA^n_{ab}$ via
\begin{align}
	\mathcal{R}^{n}_{abcd} = k_{[a} \calA^n_{b][c} k_{d]}  + i \big[( \nabla_{[a} \calA^{n-1}_{b][c}) k_{d]} + (\nabla_{[c} \calA^{n-1}_{d][a}) k_{b]} -(\nabla_{[c} k_{[a} ) \calA^{n-1}_{b]d]} \big]
	\nonumber
	\\
	~ 	+ \nabla_{[a} \nabla_{[c} \calA^{n-2}_{d]b]} - \frac{1}{2} R_{ab[c}{}^{f} \calA^{n-2}_{d]f} .
	\label{RcoeffDef}
\end{align}
Only the first term survives in the $n=0$ geometric-optics limit, and it follows immediately that $\mathcal{R}^0_{abcd}$ is trace-free and of Petrov type N with repeated principal null direction $k^a$.

The constraints \eqref{hGauge} and \eqref{hTransport} on the $\calA^n_{ab}$ imply a number of constraints on the curvature amplitudes $\mathcal{R}_{abcd}^n$. As a direct expression of the vacuum Einstein equation, the traces of the first two curvature coefficients must vanish. Beyond this,
\begin{equation}
	\mathcal{R}_{acbd}^n g^{cd} = - \tfrac{1}{2} (R_{a}{}^{c}{}_{b}{}^{d} \calA_{cd}^{n-2} + \Lambda \calA^{n-2}_{ab}).
	\label{Rtrace}
\end{equation}
Moreover, the algebraic Bianchi identity is preserved at each order:
\begin{equation}
	\mathcal{R}^n_{[abc]d} = 0.
	\label{BianchiAlg}
\end{equation}
The differential Bianchi identity is more complicated. It is convenient to first define the amplitudes
\begin{equation}
	\Gamma^n_{cab} \equiv k_{(a} \calA_{b)c}^n - \tfrac{1}{2} k_c \calA^n_{ab} + i ( \nabla_{(a} \calA^{n-1}_{b)c} - \tfrac{1}{2} \nabla_c \calA^{n-1}_{ab} ) ,
\end{equation}
which determine the perturbed first-order connection via
\begin{equation}
	\Gamma^c_{ab} = - i \omega e^{i \omega \varphi} \sum_{n=0}^\infty \omega^{-n} g^{cd} \Gamma_{dab}^n.
\end{equation}
In terms of this, the covariant derivative of a vector field $v^a$ in the perturbed metric would be $\nabla_b v^a + \Gamma^a_{bc} v^c$. Moreover, the $\Gamma^n_{cab}$ may be used to reduce the differential Bianchi identity to 
\begin{equation}
	k_{[a} \mathcal{R}^n_{bc]df} + i \nabla_{[a} \mathcal{R}^{n-1}_{bc]df} = \tfrac{1}{2} (\Gamma^{n-2}_{ed[a} R_{bc]f}{}^{e} - \Gamma^{n-2}_{ef[a} R_{bc]d}{}^{e}). 
	\label{BianchiDiff}
\end{equation}
As one application, contracting the indices $c$ and $f$ in this expression while employing \eqref{Rtrace} allows $\mathcal{R}^n_{abcd} k^d$ to be written in terms of lower-order curvature coefficients. Separately, the $\mathcal{R}^n_{abcd}$ may also be shown to satisfy a number of transport equations. These have a rather  complicated form in general, but reduce to $L \mathcal{R}^0_{abcd} = 0$ in the $n=0$ case.

\subsection{Polarization}
\label{Sect:GravPol}

Gravitational wave polarization may be understood at leading order by factorizing $\calA^0_{ab}$ using a scalar amplitude $\calA_0$ which satisfies $L \calA_0 = 0$. In terms of this, a polarization tensor $e_{ab} = e_{(ab)}$ may be introduced such that
\begin{equation}
	\calA_{ab}^0 = \calA_0 e_{ab}.
	\label{AeSplitGrav}
\end{equation}
It follows from \eqref{hTransport} that $e_{ab}$ must be parallel transported along the null rays associated with the geometric-optics approximation:
\begin{equation}
	k \cdot \nabla e_{ab} = 0.
\end{equation}
The scalars $e^{ab} e_{ab}$, $e^{ab} \bar{e}_{ab}$, and $e^{a}{}_{a}$ are thus constant along each ray. Moreover, it follows from \eqref{hGauge} that
\begin{equation}
	e_{ab} k^b = \tfrac{1}{2}k_a e^{b}{}_{b} . 
	\label{eConstrainGW}
\end{equation}
Adopting the parallel-transported null tetrad \eqref{tetrad}, the most general polarization tensor which satisfies these constraints is
\begin{equation}
	e_{ab} = e_+ m_a m_b + e_- \bar{m}_a \bar{m}_b + k{}_{(a} \chi_{b)},
	\label{eExpandgrav}
\end{equation}
where the $e_\pm$ are constant along rays and $\chi_a$ is parallel transported but otherwise arbitrary. This is closely analogous to the electromagnetic expansion \eqref{eExpand}.

Following \eqref{RcoeffDef}, it is natural to say that it is really not $e_{ab}$, but rather $k_{[a} e_{b][c} k_{d]} \propto \mathcal{R}_{abcd}^0$ which acts as a gauge-invariant polarization tensor at leading order. This is unaffected by\footnote{While $\chi_a$ cannot affect the leading-order curvature---which implies also that the trace of $e_{ab}$ cannot affect it---these statements do not necessarily apply at higher orders. See the example in \ref{Sect:kPolgrav}.} the $\chi_a$ in \eqref{eExpandgrav}, implying that it is only the $e_\pm$ coefficients which contribute to $\mathcal{R}_{abcd}^0$. These describe the two polarization states of the gravitational wave in the circularly-polarized basis $\{ m_a m_b, \bar{m}_a \bar{m}_b \}$. If $\mathcal{R}^0_{abcd} \neq 0$, there is no loss of generality in normalizing such that
\begin{equation}
	e_{ab} \bar{e}^{ab} = |e_+|^2 + |e_-|^2 = 1.
	\label{e2grav}
\end{equation}
This is assumed below unless otherwise noted, from which it follows that the characterizations \eqref{polDefs} of linearly- and circularly-polarized electromagnetic waves are unchanged for gravitational waves.

\subsection{Newman-Penrose scalars}
\label{Sect:NPgrav}

As in the electromagnetic context, it can be useful to decompose the curvature perturbation into components with respect to the tetrad \eqref{tetrad}. It is convenient in particular to decompose the trace-free component $\delta C_{abcd}$ of the curvature perturbation $\delta R_{abcd} = \Re \delta r_{abcd}$. Real trace-free tensors with Riemann-type symmetries are known to be completely characterized by the five complex Newman-Penrose scalars \cite{ExactSolns, FrolovRev, Chandra}
\begin{equation}
\label{NPgravDef}
\begin{gathered}
	\delta \Psi_0 \equiv \delta C_{abcd} k^a m^b k^c m^d, \qquad \delta \Psi_1 \equiv \delta C_{abcd} k^a n^b k^c m^d,
	\\
	\delta  \Psi_2 \equiv \delta C_{abcd} k^a m^b \bar{m}^c n^d, \qquad \delta \Psi_3 \equiv \delta C_{abcd} k^a n^b \bar{m}^c n^d ,
	\\
	\delta  \Psi_4 \equiv \delta C_{abcd} n^a \bar{m}^b n^c \bar{m}^d .
\end{gathered}
\end{equation}
An analog of \eqref{FNP}, in which $\delta C_{abcd}$ is expressed in terms of the $\delta \Psi_i$ and the tetrad, may be found in Ch. 1, Eq. (298) of \cite{Chandra}.

We now derive how these scalars scale with different powers of $\omega^{-1}$, establishing peeling results analogous to the electromagnetic scalings \eqref{Peeling0} and \eqref{Peeling}. First note that at geometric-optics order, all Newman-Penrose scalars except for $\delta \Psi_4$ vanish in the given tetrad. The leading-order curvature is therefore characterized entirely by
\begin{align}
	\delta \Psi_4 = \frac{1}{2} \omega^2 \bar{m}^a \bar{m}^b \Re ( \calA^0_{ab} e^{i \omega \varphi} ) + \mathcal{O}(\omega). 
	\label{Psi4}
\end{align} 
Substituting \eqref{AeSplitGrav} and \eqref{eExpandgrav} into this expression shows that in terms of the polarization components $e_\pm$ and the corrected scalar phase $\hat{\varphi}$ defined by \eqref{phaseCorrect},
\begin{equation}
	\delta \Psi_4 = \frac{1}{4} \omega^2 |\calA_0| ( e_+ e^{i \omega \hat{\varphi} } + \bar{e}_- e^{-i\omega \hat{\varphi} } ) + \mathcal{O}(\omega).
	\label{Psi4Pol}
\end{equation}
This is very similar to the electromagnetic scalar $\Phi_2$ given by \eqref{Phi2pol}. Its magnitude is
\begin{equation}
	|\delta \Psi_4 | = \frac{1}{4} \omega^2 |\calA_0| \big[ 1 + 2 \Re \big(e_+ e_- e^{i \omega \hat{\varphi}} \big) \big]^{1/2} + \mathcal{O}(\omega),
	\label{Psi4Mag}
\end{equation}
from which it may be seen that rapid oscillations vanish if a gravitational wave is circularly polarized.

Continuing, $\delta \Psi_3$ first appears at one order beyond geometric optics. It has the form
\begin{align}
	\delta \Psi_3 =  \frac{1}{4} \omega \bar{m}^a \Im \Big\{ \Big[ 2\nabla^b \calA^0_{ab} + \big( \delta^d_a \delta^f_b \delta^e_c + 2 \delta^d_c \delta^f_b  \delta^e_a - 4 \delta^d_a \delta^f_c \delta^e_b \big)
	\nonumber
	\\
	~ \times n^b  \calA^0_{df} \nabla_e k^c  \Big] e^{i \omega \varphi} \Big\} + \mathcal{O}(\omega^0),
\end{align}
where the latter group of terms project out any dependence on $\chi_a$ in the expansion \eqref{eExpandgrav} for $e_{ab}$. This is qualitatively similar to the electromagnetic scalar $\Phi_1$ as given by \eqref{Phi1}. Next, $\delta \Psi_2$ appears at the same order as $\delta \Psi_3$ but is significantly simpler: In terms of the shear \eqref{shearDef},
\begin{equation}
\	\delta \Psi_2 = \frac{1}{2} \omega \sigma \bar{m}^a \bar{m}^b \Im (\calA^0_{ab} e^{i\omega \varphi}) + \mathcal{O}(\omega^0).
	\label{Psi2}
\end{equation}
An explicit expression for the first non-vanishing term in $\delta \Psi_1$ is long and is omitted here. However, we do note it arises at two orders beyond geometric optics. The final Newman-Penrose scalar $\delta \Psi_0$ also arises at two orders beyond geometric optics, and is simply
\begin{equation}
	\delta \Psi_0 =- \frac{3}{2} \sigma^2 \bar{m}^a \bar{m}^b \Re ( \calA^0_{ab} e^{i \omega \varphi}) + \mathcal{O}(\omega^{-1}).
	\label{Psi0}
\end{equation}
Comparing with \eqref{Psi4}, this can also be written as $\delta \Psi_0 = -3 (\sigma/\omega)^2 \delta \Psi_4 + \mathcal{O}(\omega^{-1})$.

Without any restrictions on the nature of the ray congruence tangent to $k^a$, this discussion implies that in general,
\begin{equation}
	\delta \Psi_4 = \mathcal{O}(\omega^2), \qquad \delta \Psi_3, \delta \Psi_2 = \mathcal{O}(\omega^1), \qquad \delta \Psi_1, \delta \Psi_0 = \mathcal{O}(\omega^0).
	\label{NPscaleGrav}
\end{equation}
As in the electromagnetic case, a result more reminiscent of the usual gravitational peeling theorems---formulated in powers of inverse distance \cite{PenrosePeeling} instead of inverse frequency---arises when $\sigma = 0$; the shear-free case may be summarized by
\begin{equation}
\begin{gathered}
	\delta \Psi_i = \mathcal{O}(\omega^{i-2}), \qquad i = 2,3,4,
	\\
	\delta \Psi_1 = \mathcal{O}(\omega^0) , \qquad \delta \Psi_0 = \mathcal{O}(\omega^{-1}) .
\end{gathered}
\end{equation}
However, the calculations carried out here are not sufficient to decide if the results in the second line may be sharpened.

\subsection{Principal null directions}

In Sect. \ref{Sect:EMPND}, we considered the principal null directions of the electromagnetic field as generalized ``propagation directions.'' The same may be done for gravitational waves, in which case the principal null directions of interest are those associated with $\delta C_{abcd}$. In particular, consider those $k'^a$ which are null with respect to $g_{ab}$ and which satisfy
\begin{equation}
	k'_{[a} \delta C_{b]cd[f} k'_{e]} k'^c k'^d = 0.
	\label{PNDgrav}
\end{equation}
Note that this is somewhat different from asking for principal null directions associated with the spacetime as a whole. Nontrivial backgrounds generically admit their own such directions, even in the absence of any perturbation at all. These are not interesting as descriptions for the overlying gravitational waves, although such a statement clearly relies on a reliable way of distinguishing the background and perturbed geometries. This distinction is aided by the aforementioned gauge-invariance of the curvature perturbation at lower orders, although it is nontrivial in general. 

Our approach to solving \eqref{PNDgrav} is similar to that used for its electromagnetic analog \eqref{PNDdef}. Again adopting a primed tetrad with the form \eqref{tetradPrime}, it may be shown that $k'^a$ is a real principal null vector if and only if $\delta \Psi'_0 = \delta C_{abcd} k'^a m'^b k'^c m'^d = 0$ \cite{Chandra}. This may be used as a criterion with which to find $k'^a$, starting with the unprimed tetrad \eqref{tetrad} and then using \eqref{tetradRotate} to rotate $k^a$ into an appropriate solution. In terms of the complex $z$ which parametrizes that rotation,
\begin{equation}
	\delta \Psi'_0 = \delta \Psi_0 + 4 z \delta \Psi_1 + 6 z^2 \delta \Psi_2 + 4 z^3 \delta \Psi_3 + z^4 \delta \Psi_4 = 0.
	\label{PNDeqnGrav}
\end{equation}
Solving this equation for $z$ recovers the principal null congruences. It is a quartic equation, so there are four such congruences in general.

In general, solutions for $z$ oscillate with the same frequency as $h_{ab}$. Moreover, $\delta \Psi_4$ oscillates through zero at this frequency for linearly-polarized waves. These complications may be avoided by restricting considerations to circularly-polarized waves. Choosing $m_a$  such that $e_+ = 1$ and $e_- = 0$ in \eqref{eExpand}, it follows from \eqref{Psi4Pol}, \eqref{Psi2}, and \eqref{Psi0} that
\begin{equation}
\begin{gathered}
	\delta \Psi_4 = \frac{1}{4} \omega^2 \calA_0 e^{i \omega \varphi} + \mathcal{O}(\omega), \qquad \delta \Psi_2 = - \frac{1}{4} i \omega \sigma \calA_0 e^{i \omega \varphi} + \mathcal{O}(\omega^0),	\\
	\delta \Psi_0 = - \frac{3}{4} \sigma^2 \calA_0 e^{i \omega \varphi} + \mathcal{O}(\omega^{-1}),
\end{gathered}
\end{equation}
so $3 \delta \Psi_2^2 - \delta \Psi_0 \delta \Psi_4 = \mathcal{O}(\omega)$. Substituting these expressions into \eqref{PNDeqnGrav} and solving to lowest nontrivial order,
\begin{equation}
	z = \pm \left[ i (3 \pm \sqrt{6}) (\sigma/\omega) \right]^{1/2}  + \mathcal{O}(\omega^{-1}).
\end{equation}
Eq. \eqref{tetradRotate} thus implies that the principal null directions are given by
\begin{equation}
	k'^a = k^a \pm 2 \Re \left\{ \left[  - i (3 \pm \sqrt{6} ) \bar{\sigma}/\omega \right]^{1/2} m^a \right\} + \mathcal{O}(\omega^{-1}) 
	\label{pertPNDgrav}
\end{equation}
for a circularly-polarized gravitational wave with polarization $e_{ab} = m_a m_b + k_{(a} \chi_{b)}$. Like its electromagnetic analog \eqref{pertPND}, the first corrections here depend on the shear of the underlying congruence and scale like $\omega^{-1/2}$, not $\omega^{-1}$. If $\sigma \neq 0$, the single leading-order principal null vector splits into four vectors already at this order. There are thus four effective ``propagation directions'' associated even with a circularly-polarized gravitational wave. If $\sigma = 0$, the principal null directions differ from $k^a$ instead by terms of order $\omega^{-1}$. Four directions generically appear at this order, although there are exceptions where two or more directions remain degenerate.

\subsection{Other observables}

In the electromagnetic context, propagation directions were associated in Sect \ref{Sect:EMT} not only with the principal null directions of $F_{ab}$, but also with its averaged stress-energy tensor. It is much less clear that a similar calculation would be physically interesting for gravitational waves. While Isaacson's stress-energy tensor \cite{Isaacson2, MTW, Burnett} may be interpreted as explaining the averaged gravitational backreaction due to a high-frequency gravitational wave, existing and proposed methods of gravitational-wave detection do not directly probe this; the perturbed curvature is instead the most natural observable. Moreover, it is not clear that Isaacson's stress-energy is meaningful in a regime where expansions are performed beyond geometric optics while nonlinearities in Einstein's equation are ignored; it is derived assuming specific relations between a wave's amplitude, its frequency, and an external lengthscale---relations which are not necessarily appropriate to the finite-frequency discussions considered here.

An alternative approach which avoids many of these difficulties would be to consider a ``superenergy tensor'' associated with a high-frequency gravitational wave\footnote{Superenergy tensors may also be associated with non-gravitational fields; see, e.g., \cite{Senovilla2000} and references therein.}. The prototypical example is the Bel-Robinson tensor, which is a rank-4, divergence-free tensor field which is quadratic in the Weyl tensor. Its definition does not depend on any type of approximation or averaging procedure. If a high-frequency expansion is nevertheless applied to the perturbed Bel-Robinson tensor, the leading-order result may be shown to be
\begin{align}
	\langle \delta T_{abcd} \rangle & \equiv \langle \delta C_{aecf} \delta C_{b}{}^{e}{}_{d}{}^f + \delta C_{aecf}^* \delta C^*_{b}{}^{e}{}_{d}{}^f \rangle
	\nonumber
	\\
	& = \frac{ 1 }{ 16 } \omega^4 \|\calA_0 \|^2 k_a k_b k_c k_d + \mathcal{O}(\omega^3),
	\label{Bel}
\end{align}
where 
\begin{equation}
	\| \calA_0 \|^2 \equiv ( g^{ac} g^{bd} - \frac{1}{2} g^{ab} g^{cd}) \calA^0_{ab} \bar{\calA}^0_{cd}
	\label{gravNorm}
\end{equation}
is a norm which eliminates any dependence on the trace of $\calA^0_{ab}$. The tensorial structure here is as expected for a Petrov type-N Weyl tensor; the single relevant propagation direction is $k^a$ and the amplitude is given by $\langle | \delta \Psi_4 |^2 \rangle$. While it would be interesting to expand $\langle \delta T^{abcd} \rangle$ to higher orders in $\omega^{-1}$ and to interpret the resulting corrections in terms of the principal null vectors \eqref{pertPNDgrav}, this is left for later work.

Even at leading order, \eqref{Bel} is not uninteresting. Noting that the Bel-Robinson tensor has units of $\mbox{(length)}^{-4}$, which differs from the $\mbox{(length)}^{-2}$ associated with an ordinary stress-energy tensor, there has been some uncertainty regarding its physical interpretation (independently of any particular approximation). One possibility which has been proposed is that a rank-2 square root of the Bel-Robinson tensor may serve as a kind of gravitational stress-energy tensor \cite{Bonilla1997}; this has been used to propose a notion of gravitational entropy \cite{EllisEntropy} and also to discuss interactions between material and gravitational fields \cite{EllisEnergy}. While square roots do not exist for all geometries, there are no difficulties at high frequencies; inspection of \eqref{Bel} shows that
\begin{equation}
	\frac{ 1 }{ 4 }  \omega^2 \|\calA_0\| k_a k_b + \mathcal{O}(\omega)
	\label{Belroot}
\end{equation}
is such a root. This has the same tensorial structure as the geometric-optics stress-energy tensors associated with scalar and electromagnetic fields, as has been noted previously for Bel-Robinson tensors  associated with type-N curvature tensors \cite{Bonilla1997, EllisEntropy, EllisEnergy}. However, the coefficient here does not support the analogy: Reasonable stress-energy tensors should scale like the square of the field amplitude, not the amplitude itself. It thus appears to be dubious to interpret the square root of the Bel-Robinson tensor as a kind of stress-energy tensor. Rather, the extra units of $\mbox{(length)}^{-2}$ in $\delta T^{abcd}$ may be better thought of as an inverse area. Although the reasoning is different, the same conclusion is reached in \cite{Senovilla2000}.

A kind of propagation direction which is not associated directly with the principal null directions, an effective stress-energy tensor, or the Bel-Robinson tensor is that associated with a generalized area-intensity law. First note that the current $J_0^a$ defined by \eqref{J0scalar} remains conserved as-is in the gravitational theory. However, the norm $|\cdot|$ which appears there is perhaps inappropriate in light of \eqref{Bel}. A better choice would be to define
\begin{equation}
	J^a_{0'} \equiv \| \calA_0 \|^2 k^a,
	\label{J0prime}
\end{equation}
and it may be verified that this is conserved as well. Continuing to subleading order, the electromagnetic $J^a_1$ given by \eqref{J1em} generalizes straightforwardly; its gravitational counterpart is
\begin{equation}
	J^a_1 \equiv 2 \Re ( \calA_0^{bc} \bar{\calA}^1_{bc} ) k^a - |\calA_0|^2 \big[ \nabla^a \arg \calA_0 + i e^{bc} \nabla^a \bar{e}_{bc} \big] .
\end{equation}
However, this is better associated with the norm $| \cdot |$ instead of $\| \cdot \|$. Another vector field which matches better with the latter norm is
\begin{align}
	J^a_{1'} \equiv 2 ( g^{bd} g^{cf} - \tfrac{1}{2} g^{bc} g^{df} ) \Re ( \calA^0_{bc} \bar{\calA}^1_{df}  ) k^a - |\calA_0|^2 \big[ (1 - \tfrac{1}{2} | e^{b}{}_{b} |^2 ) \nabla^a \arg \calA_0 
	\nonumber
	\\
	~ + i ( e^{bc} \nabla^a \bar{e}_{bc} - \tfrac{1}{2} e^{b}{}_{b} \nabla^a \bar{e}^{c}{}_{c} ) \big] ,
\end{align}
and this too is conserved. Either the unprimed or primed currents may be added together to find conservation laws which express area-intensity relations at subleading order. In the primed case (which is more complicated but likely to be more physical), the vector
\begin{align}
	J_{0'}^a + \omega^{-1} J_{1'}^a = \| \calA_0 + \omega^{-1} \calA_1 \|^2 \big\{ k^a - \omega^{-1} |\calA_0|^2 \big[  (1 - \tfrac{1}{2} | e^{b}{}_{b} |^2 ) \nabla^a \arg \calA_0  
	\nonumber
	\\
	~ + i ( e^{bc} \nabla^a \bar{e}_{bc} - \tfrac{1}{2} e^{b}{}_{b} \nabla^a \bar{e}^{c}{}_{c} ) \big]\big\} + \mathcal{O}(\omega^{-2})
	\label{areaLawGrav}
\end{align}
is conserved. The vector on the right-hand side here is real and null, and may be interpreted as a kind of correction to the propagation direction. The cross-sectional areas of the congruence tangent to it control variations in $\| \calA_0 + \omega^{-1} \calA_1 \|^2$.

Observables which are more directly physical may be obtained by considering the motions of freely-falling test particles. If a freely-falling observer with 4-velocity $u^a$ measures the separation $\xi^a$ of a nearby test particle, the geodesic deviation equation implies that the relative acceleration of that particle involves $\delta R^{a}{}_{bcd} u^b u^c \xi^d$. At leading order, the high-frequency contribution to this which is implied by \eqref{rAnsatz} and \eqref{RcoeffDef} is orthogonal both to $u^a$ and to $k^a$, as expected from the transverse nature of gravitational radiation in general relativity. A kind of longitudinal acceleration may nevertheless arise when expanding beyond geometric optics. To see this in a simple context, note that if $\xi^a$ has a component proportional to a projection of $k^a$ into the observer's rest frame, the longitudinal acceleration involves
\begin{align}
	k^a (\delta r_{abcd} u^b u^c k^d) = \frac{1}{2} i (\omega_o^2/\omega) |\calA_0| (e_+ \sigma e^{i \omega \hat{\varphi} } + e_- \bar{\sigma} e^{-i \omega \hat{\varphi} } )  + \mathcal{O}(\omega^0),
\end{align}
where $\hat{\varphi}$ is the corrected phase \eqref{phaseCorrect} and $\omega_o$ is the measured frequency \eqref{omegaO}. This depends on the shear $\sigma$. 

The last observables we consider for a gravitational wave are scalars constructed from $\delta C_{abcd}$. Four such scalars may be locally constructed without differentiating or introducing an external frame. These are either quadratic
\begin{equation}
	\delta C^{abcd} \delta C_{abcd} , \qquad \delta C^{abcd} \delta C^*_{abcd},
\end{equation}
or cubic,
\begin{equation}
	\delta C^{abcd} \delta C_{cd ef} \delta C^{ef}{}_{ab}, \qquad \delta C^{abcd} \delta C_{cd}{}^{ef} \delta C^{*}_{efab},
\end{equation}
where $\delta C^*_{abcd} = \frac{1}{2} \epsilon_{cd}{}^{ef} \delta C_{abef}$ denotes the right dual of the perturbed Weyl tensor (although the right and left duals are equal here). Computing averages using the prescription given by \eqref{Sbilinear} and \eqref{avDef} shows that both quadratic scalars are suppressed by at least three powers of $\omega$ relative to generic components of $\langle \delta C_{abcd} \delta C_{efgh} \rangle = \mathcal{O}(\omega^4)$:
\begin{align}
	\langle \delta C^{abcd} \delta C_{abcd} \rangle = \mathcal{O}(\omega), \qquad \langle \delta C^{abcd} {}^*\delta C_{abcd} \rangle  = \mathcal{O}(\omega).
\end{align}
All cubic scalars vanish under averaging.

\section{Relating different types of high-frequency fields}

To summarize the starting points for the above discussions, high-frequency approximations for scalar, electromagnetic, and gravitational waves were found to be governed by the eikonal equation \eqref{eikonal} and the transport equations
\begin{equation}
	L \calA_B^n = - i \mathcal{D} \calA_B^{n-1},
	\label{transGen}
\end{equation}
where $L$ is the operator \eqref{transOp}, $B$ is a multi-index appropriate to the field under consideration, and $\mathcal{D} = \Box + \ldots$ is the hyperbolic operator associated with the appropriate field equation; Eqs. \eqref{phi0trans}, \eqref{phiNtrans}, \eqref{xPortEM}, and \eqref{hTransport} are all in this form. In the electromagnetic and gravitational cases, the amplitudes must also satisfy the algebraic constraints \eqref{aGauge} and \eqref{hGauge}. At this level, it may appear that there is very little difference between the various types of fields considered here. We now discuss to what extent differences do exist, and also when similarities may be exploited to simplify calculations. When, for example, does solving an effective scalar (or electromagnetic) problem suffice to understand a problem which is physically electromagnetic (or gravitational)?

\subsection{Leading-order amplitudes}

The clearest cases in which such simplifications arise are those which depend only locally\footnote{It follows from \eqref{transGen} that up to homogeneous solutions, all higher-order amplitudes may be viewed as functionals of the $n=0$ amplitudes. However, these functionals are nonlocal in general; they involve integrals along null geodesics. Nevertheless, there are many cases in which the dependence relevant to a particular observable at a particular order reduces to a local function of the leading-order amplitude and a finite number of its derivatives.} on the $n=0$ amplitudes. These leading-order amplitudes locally determine all of geometric optics, but also much more than this: All approximations for the Newman-Penrose scalars given in Sects. \ref{Sect:NPEM} and \ref{Sect:NPgrav} are written solely in terms of the $n=0$ amplitudes and the geometric-optics propagation direction $k^a$, even though it is only $\Phi_2$ and $\delta \Psi_4$ which characterize geometric-optics fields. Similarly, the variously-defined corrections \eqref{Kscalar},  \eqref{pertPND}, \eqref{kDefem}, \eqref{pertPNDgrav}, and \eqref{areaLawGrav} to $k^a$ are locally written using only leading-order quantities. An understanding for how $\calA_0$, $\calA^0_a$, and $\calA_{ab}^0$ relate to one another provides new insights into these quantities and others.

The first such relation is associated with the fact that solving a scalar problem automatically solves aspects of electromagnetic and gravitational problems, in the sense that if a leading-order scalar amplitude $\calA_0$ is known, \eqref{AeSplit}, \eqref{e2}, \eqref{AeSplitGrav}, and \eqref{e2grav} imply that its square $|\calA_0|^2 = \calA_0 \bar{\calA}_0$ also determines the squares $g^{ab} \calA^0_a \bar{A}_b^0$ and
$g^{ab} g^{cd} \calA^0_{ab} \bar{A}_{cd}^0$ of leading-order electromagnetic and gravitational amplitudes. These quantities appear in the averaged stress-energy and superenergy tensors at leading order. In fact, the entirety of the averaged electromagnetic stress-energy tensor may be determined at this order by solving a scalar problem; see \eqref{TemFact}. This is also true for the leading-order average \eqref{Bel} of the Bel-Robinson tensor in the gravitational case, at least if\footnote{This caveat is not essential. It may be removed by modifying the normalization condition \eqref{e2grav}.} $e^{a}{}_{a} = 0$. In geometric optics, observables such as the averaged energy and momentum densities and the propagation direction may thus be understood purely by solving scalar problems; the additional complexities of the electromagnetic and gravitational amplitudes do not affect these quantities at leading order.

Of course, not all leading-order observables may be understood so simply: Even within geometric optics, there are fundamental differences between $\nabla_a \Psi$, $F_{ab}$, and $\delta R_{abcd}$. An infinite variety of electromagnetic fields with distinct field strengths may, for example, be associated with the same scalar problem and the same leading-order $\langle T_{ab} \rangle$. Nevertheless, if a scalar amplitude $\calA_0$ is supplemented with a parallel-transported transverse polarization $e_a$, that amplitude trivially maps to an electromagnetic one via $\calA^0_a = \calA_0 e_a$. Comparison of \eqref{gradPsi}, \eqref{Fansatz}, and \eqref{FnDef} shows that in addition, field strengths are related via
\begin{equation}
	f_{ab} = - 2 e_{[a} \nabla_{b]} \psi + \mathcal{O}(\omega^0)
\end{equation}
under this mapping, where the error term here is at one order beyond geometric optics.

If an electromagnetic amplitude is instead used to construct a gravitational amplitude, the gravitational polarization tensor may be constructed entirely from the electromagnetic polarization; no supplementary information is required. Suppose in fact that there are two known electromagnetic amplitudes, $\calA_a^0 = \calA_0 e_a$ and $\calA_a'^0 = \calA_0 e_a'$, which may have different polarizations. Then,
\begin{equation}
	\calA^0_{ab} = \calA_0 [ e_{(a} e'_{b)} -\tfrac{1}{2} g_{ab} ( e \cdot e') ]
	\label{doubleMetric}
\end{equation}
satisfies the gravitational constraint and transport equations \eqref{hGauge} and \eqref{hTransport} and is therefore a valid gravitational amplitude. The normalization condition \eqref{e2grav} is not necessarily preserved by this mapping [assuming that $e_a$ and $e'_a$ satisfy \eqref{e2}], although this is easily restored by rescaling $\calA_0$ in \eqref{doubleMetric}. It is however simpler not to do this. Then the gravitational norm \eqref{gravNorm} may be computed, assuming the electromagnetic normalization \eqref{e2}. Using $|\calA_0|^2$ to denote the square of the scalar amplitude,
\begin{align}
	\|\calA_0\|^2 &= \frac{1}{2} |\calA_0|^2 \big[ 1 + \Re ( e \cdot \bar{e}' ) - |e \cdot e'|^2 \big] .
	\label{gravNormDoub}
\end{align}
If both electromagnetic waves are identical and circularly polarized, it follows from this that $\|\calA_0\|^2 = |\calA_0|^2$. If both waves are circularly polarized but with opposite helicities, $\|\calA_0\|^2 = 0$.

These results can be understood more generally by resolving the various polarizations into the circularly-polarized bases $\{ m_a , \bar{m}_a \}$ and $\{ m_a m_b, \bar{m}_a \bar{m}_b \}$ using \eqref{eExpand} and \eqref{eExpandgrav}. It then follows from \eqref{doubleMetric} that the electromagnetic and gravitational polarization states are related via
\begin{equation}
	e_\pm^g = e_\pm e_\pm' 
	\label{eTransform}
\end{equation}
and
\begin{align}
	\chi^a =\chi \chi' k^a +  ( \chi e_+' + e_+ \chi' ) m^a + (\chi e_-' + \chi' e_-) \bar{m}^a 
	 + (e_+ e_-' + e_+' e_-) n^a ,
\end{align}
where the ``$g$'' superscript has been inserted to distinguish the gravitational components. The term involving $\chi^a$ is irrelevant at leading order, so there is a sense in which the gravitational polarization components are simply products of the underlying electromagnetic components: If at least one of the electromagnetic waves used here is circularly polarized, so is the resulting gravitational wave. If both waves are circularly polarized but with opposite helicities, the associated gravitational wave vanishes at leading order---as already expected from \eqref{gravNormDoub}.

Regardless of polarization, the curvature perturbation $\delta r_{abcd}$ associated with the amplitude \eqref{doubleMetric} may be computed from the electromagnetic field strengths $f_{ab}$ and $f'_{ab}$. Using \eqref{FnDef} and \eqref{RcoeffDef},
\begin{equation}
	\mathcal{R}_{abcd}^0 = \calA_0^{-1} \big[ 2g^{fh} \mathcal{F}^0_{f[a} g_{b][c} \mathcal{F}'^0_{d]h}  - \tfrac{1}{2} \big( \mathcal{F}^0_{ab} \mathcal{F}'^0_{cd} + \mathcal{F}'^0_{ab} \mathcal{F}^0_{cd} \big) \big],
\end{equation}
where $\mathcal{F}^0_{ab}$ and $\mathcal{R}^0_{abcd}$ are the leading terms in the expansions \eqref{Fansatz} and \eqref{rAnsatz}. Up to a scalar factor, this shows that $\mathcal{R}^0_{abcd}$ is the trace-free symmetrized product of $\mathcal{F}_{ab}^0$ and $\mathcal{F}'^0_{ab}$. Alternatively, $\mathcal{F}_{ab}^0$ may be viewed as a square root of $\mathcal{R}^0_{abcd}$ when the primed and unprimed fields coincide.

The correspondence \eqref{doubleMetric} might be interpreted as a kind of classical double copy. Two electromagnetic solutions are ``copied into,'' or ``squared'' to produce a single gravitational solution. This language is borrowed from quantum field theory, where it is known that under certain conditions, gravitational scattering amplitudes look like gauge-theory amplitudes ``squared'' \cite{Bern2010}. Such results have inspired significant discussion of classical analogs in which solutions to gauge-theory equations generate solutions to gravitational equations (often coupled to non-gravitational fields); see \cite{DoubleCopy2,DoubleCopy} and references therein. Much of the discussion on the classical gravitational side has been confined to metrics of Kerr-Schild type, i.e. those in which a background is deformed by adding to it a term with the form $V k_a k_b$, where $k_a$ is null. The double copy given by \eqref{doubleMetric} includes at least the $pp$-waves in this class; see \ref{Sect:kPolgrav}. However, the correspondence given here between electromagnetic and gravitational solutions in fact holds in general in geometric optics. In special cases where geometric optics is exact---as for $pp$-waves---it extends to exact solutions. However, given that the non-classical double copy results are associated with scattering, it is perhaps reasonable to expect a classical analog to be generic mainly in the geometric-optics regime which is so central to scattering calculations.

\subsection{Higher-order amplitudes}
\label{Sect:HighRel}

Even though it is simple to relate the $n=0$ amplitudes associated with scalar, electromagnetic, and gravitational waves, these relations do not necessarily survive at higher orders. One exception---where well-defined higher-order relations \textit{can} be derived between different theories---involves scalar theories with different masses $\mu$ or curvature couplings $\xi$. These possibilities behave identically at the level of the leading-order amplitudes, but not more generally. Concentrating on the $n=1$ case, suppose that one leading-order amplitude $\calA_0$ is known and that this is used to determine corrected amplitudes associated with two different types of scalar field. Letting $(\mu,\xi)$ and $(\mu',\xi')$ be the parameters which characterize those fields, \eqref{phiNtrans} implies their corrected amplitudes must satisfy
\begin{equation}
	L ( \mathcal{A}'_1 - \mathcal{A}_1 ) = i \big[ ( \xi' - \xi ) R + \mu'^2 - \mu^2 \big] \calA_0.
	\label{LAdiff}
\end{equation}
This can be solved by introducing an affine distance $r$ which satisfies \eqref{rDef}. If $\calA_1$ and $\calA'_1$ are assumed to coincide on a hypersurface $r = r_0$, where $k \cdot \nabla r_0 = 0$, 
\begin{equation}
	\calA_1' = \calA_1 + \frac{i}{2} \calA_0 \left[ ( \xi'- \xi) \int_{r_0}^r R dr'  + ( \mu'^2 - \mu^2 ) (r - r_0) \right]   .
	\label{scalarA1}
\end{equation}
The integral here is along a ray which connects the $r=r_0$ hypersurface to the point at which the amplitude is evaluated. Regardless, it indicates that the subleading amplitudes for fields with different masses grow with $r$. Amplitudes associated with different curvature couplings instead grow with the integral of the Ricci scalar. While these terms may be large for radiation emitted by distant sources, their primary effect is to shift the phase of $\psi'$ relative to that of $\psi$: Differing masses result in the phase shift $\frac{1}{2} ( \mu'^2 - \mu^2 ) (r - r_0)/\omega$, and if the vacuum Einstein equation \eqref{Einstein} holds, differing curvature couplings produce the additional phase difference $2 \Lambda (\xi'- \xi) (r-r_0)/\omega$. These shifts do not, however, affect intensities as given by \eqref{Psi2Av} and \eqref{Iscalar}. More generally, it follows from \eqref{TScalarFact} that the entirety of the averaged stress-energy tensor is independent of $\mu$ and $\xi$ to the orders computed here: $\langle T'_{ab} \rangle = \langle T_{ab} \rangle + \mathcal{O}(\omega^0)$. Masses and curvature couplings do affect at least the trace \eqref{TtrSc} of this tensor at the following order, although not via any differences between $\calA_1$ and $\calA'_1$.

Relations between $n=1$ amplitudes are much less clear when comparing scalar and electromagnetic quantities. Unfortunately, even if the $n=0$ amplitudes are related via \eqref{AeSplit}, no simple result appears to follow at the following order. This may be seen by noting from \eqref{xPortEM} that $\calA^1_a$ is determined by a transport equation whose right-hand side involves
\begin{equation}
	\Box \calA^0_a = e_a \Box \calA_0 + 2 \nabla^b \calA_0 \nabla_b e_a + \calA_0 \Box e_a,
\end{equation}	
the latter two terms of which are not usefully related to the $\Box \calA_0$ which determines the $\mu=\xi = 0$ scalar amplitude $\calA_1$ via \eqref{phiNtrans}. However, one might instead ask the weaker question of whether or not any relation exists between the contributions of the $n=1$ amplitudes to $\langle \Psi^2 \rangle$ and $\langle A^2 \rangle$. These quantities set the overall scales of the stress-energy tensors \eqref{TScalarFact} and \eqref{TemFact}. In the scalar case, it follows from \eqref{Psi2Av} that the relevant quantity is $\Re (\calA_0 \bar{\calA}_1)$. In the electromagnetic case, it is $\Re (\calA_0 \cdot \bar{\calA}_1)$. Assuming the normalization condition \eqref{e2}, the difference between these quantities satisfies the transport equation
\begin{align}
 (L + \nabla \cdot k) \Re (\calA_0 \cdot \bar{\calA}_1 - \calA_0 \bar{\calA}_1) = i \nabla^b ( |\calA_0|^2  \bar{e}^a \nabla_b e_a ) .
\end{align}
Scalar computations would therefore be sufficient to determine $\langle A^2 \rangle$ when the right-hand side of this equation vanishes. A similar conclusion could also have been reached by comparing the scalar and electromagnetic conserved currents $J^a_1$, as given by \eqref{J1scalar} and \eqref{J1em}. Regardless, the term $\bar{e}^a \nabla_b e_a$ which appears here vanishes if, e.g., a wave is linearly polarized, $\chi = 0$, and the polarization angle is constant in the sense that $e_+ = e^{i \theta} \bar{e}_-$ for some real constant $\theta$. In these cases, one can actually say much more than that $\langle \Psi^2 \rangle = \langle A^2 \rangle + \mathcal{O}(\omega^{-2})$. Eq. \eqref{TemSimp} and the surrounding discussion implies that all of $\langle T_{ab} \rangle$ then coincides for scalar and electromagnetic fields, at both leading and subleading orders.

\section{Discussion}

We have derived a number of general features of high-frequency scalar, electromagnetic, and gravitational waves propagating on curved background spacetimes, focusing on observables, physical intuition, and also relations between these different types of fields. However, no specific applications were considered. The purpose has been instead to set the stage for further exploration.

While it would be straightforward to use the results presented here to compute corrections to geometric optics in various scenarios, subsequent papers in this series will take a more foundational approach. Two basic questions will be addressed before considering the details associated with any specific systems: First, how do changes in the background metric affect observables? General invariance properties of the underlying equations will be shown to provide powerful tools with which to address this question. Second, we ask how the measured properties of a radiated field can be related to intrinsic properties of its source. Alternatively, how should initial data be specified for the various transport equations? Although the space of possibilities is large in general, gravitational lensing is typically concerned with compact sources. In this context, the initial data problem simplifies considerably. We shall discuss how this occurs and how the relevant data can be related to a source's intrinsic properties. 

\appendix

\section{Nature of the high-frequency approximation}

\label{App:Approx}

This appendix discusses the errors incurred by truncating the high-frequency series discussed above. As summarized following \eqref{PhiAnsatz}, truncating the scalar series at order $m$ results in a field $\Psi_m \equiv e^{i \omega \varphi} \sum_{n=0}^m \omega^{-n} \calA_n$ which approximates a solution to the Klein-Gordon equation \eqref{KleinGordon} in the sense that 
\begin{equation}
( \Box - \xi R - \mu^2 ) \Psi_m = \mathcal{O}(\omega^{-m}). 
\end{equation}
This may be verified by directly applying the transport equations \eqref{phi0trans} and \eqref{phiNtrans}. 

Analogous statements are more complicated in the electromagnetic and gravitational contexts, essentially because the gauge conditions employed here are first-order differential equations while the field equations are second order. Suppose that the expansion \eqref{Aansatz} for the vector potential is truncated at order $m$, leaving the finite series $a_a^m$. This differs from the full series $a_a$ by terms of order $\omega^{-m-1}$. Furthermore,
\begin{equation}
	\nabla^b \nabla_b a_a^m - R_{a}{}^{b} a_b^m = \mathcal{O}(\omega^{-m}), \qquad \nabla^a a_a^m = \mathcal{O}(\omega^{-m}).
	\label{Aapprox}
\end{equation}
If a similarly-truncated field strength $f_{ab}^m \equiv 2 \nabla_{[a} a_{b]}^m$ is defined, it differs from $f_{ab}$ by terms of order $\omega^{-m}$. However, \eqref{Fansatz} and \eqref{FnDef} imply that it includes some terms which are also of order $\omega^{-m}$. Although these terms might appear to be negligible, it can be important to retain them.

In particular, these terms are important when differentiating $f_{ab}^m$ to verify that the (gauge-independent) Maxwell equation $\nabla^b f_{ab} = 0$ has been satisfied to an appropriate order. Including those terms and using \eqref{Aapprox} to evaluate the left-hand side of this equation for the truncated field,
\begin{align}
	\nabla^b f_{ab}^m  &= \nabla_a (\nabla \cdot a^m) - (\Box a_a^m - R_{a}{}^{b} a_b^m)
	\nonumber
	\\
	&= - i \omega^{1-m} e^{i \omega \varphi} ( \nabla \cdot \calA_m) k_a +  \mathcal{O}(\omega^{-m}).
	\label{MaxwellError}
\end{align}
Solving the gauge-fixed equations up to terms of order $\omega^{-m}$ thus results in an ``error current'' in Maxwell's equations which may involve a (formally larger) term of order $\omega^{1-m}$. This term is however proportional to $k_a$; other components do fall off as expected in the sense that $k_{[a} \nabla^c f^m_{b]c} = \mathcal{O}(\omega^{-m})$. If the terms of order $\omega^{-m}$ in $f_{ab}^m$ were omitted, even this latter statement would fail.

Regardless, this appears to be a situation in which Maxwell's equations are not satisfied as well as one might like. This can be addressed by adding a correction term of order $\omega^{-m}$ to $f_{ab}^m$ which eliminates the error current. We do so by defining the modified vector potential
\begin{equation}
	\mathfrak{a}_a^{m} \equiv a_a^m + \omega^{-m} e^{i \omega \varphi} \alpha_m k_a,
	\label{ANprime}
\end{equation}
where the scalar correction $\alpha_m$ satisfies the transport equation
\begin{equation}
	k \cdot \nabla \alpha_m = \nabla \cdot \calA_m.
	\label{alphaDef}
\end{equation}
If this equation is solved and the result substituted into \eqref{ANprime}, the resulting vector potential generates a corrected field strength $\mathfrak{f}_{ab}^{m} = 2 \nabla_{[a} \mathfrak{a}_{b]}^{m}$ which does satisfy Maxwell's equations up to the given order: $\nabla^b \mathfrak{f}_{ab}^m = \mathcal{O}(\omega^{-m})$. 

Similar comments also apply for metric perturbations. Solving \eqref{hGauge} and \eqref{hTransport} to obtain gravitational amplitudes up to order $m$ results in a truncated $h_{ab}^m$ which solves \eqref{LorenzGrav} and \eqref{linEins} up to terms of order $\omega^{-m}$. However, applying the full linearized Einstein operator (which does not assume any particular gauge condition) to $h_{ab}^m$ results in a ``leftover'' stress-energy tensor which is of the form $\omega^{1-m}(\ldots)_{(a} k_{b)} + \mathcal{O}(\omega^{-m})$. The first term here may be eliminated by defining a corrected metric perturbation $\mathfrak{h}_{ab}^m = h_{ab}^m + \omega^{-m} e^{i \omega \varphi} \beta^m_{(a} k^{}_{b)}$, where $\beta^m_a$ satisfies a transport equation similar to \eqref{alphaDef}. 

The corrections discussed in this appendix are not used in the body of the paper. They do not arise in practice because, for any quantity which might be of interest, it is straightforward to determine how many amplitudes are needed to guarantee a particular error. The perspective taken here is not to fix a truncation at the outset and to perform every calculation using exactly those terms, but rather to adapt the number of terms as required to each individual calculation. Doing so, there is no need to introduce $\mathfrak{a}_a^m$ or $\mathfrak{h}^m_{ab}$.

%Also note that the differences $\hat{a}_a^{N} - a^N_a$ and $\hat{f}_{ab}^{N} - f_{ab}^N$ are both $\mathcal{O}(\omega^{-N})$. Below, we consider only approximations for $f_{ab}$ itself, but not its derivatives, so these corrections are not necessary. 

\section{Polarization tensors for which $e_\pm = 0$}

The majority of this paper assumes that the polarization tensors $e_a$ and $e_{ab}$ defined by \eqref{AeSplit} and \eqref{AeSplitGrav} are normalized such that $e \cdot \bar{e} = 1$. This is always possible if at least one of the $e_\pm$ in \eqref{eExpand} or \eqref{eExpandgrav} is nonzero. However, it can also be interesting to consider nonzero amplitudes for which $e_\pm = 0$. The remaining terms, controlled by $\chi$ or $\chi_a$, do not contribute to leading-order field strengths or curvatures. However, they may affect these quantities at higher orders. We now provide examples which illustrate the meanings of these kinds of polarization tensors.

\subsection{Electromagnetism}
\label{Sect:kPol}

Suppose that the leading-order amplitude $\calA_a^0$ for the Lorenz-gauge vector potential may be written as $\calA^0_a = \calA_0 k_a$, where $\calA_0$ satisfies the scalar transport equation $L \calA_0 = 0$. This is equivalent to setting $e_\pm = 0$ in \eqref{eExpand} and absorbing $\chi$ into a redefinition of $\calA_0$. It is evident from \eqref{FnDef} that $\mathcal{F}^0_{ab} = 0$.

An explicit example of this type may be constructed by considering a wave propagating in Minkowski spacetime. Let $(t,x,y,z)$ be globally-inertial coordinates and set $\varphi = t-z$, so constant-phase hypersurfaces are hyperplanes and $k^a = \partial_t + \partial_z$. It then follows from \eqref{phi0trans} and \eqref{transOp} that the scalar transport equation is solved by any function with the form $\calA_0 = \calA_0 (\varphi,x,y)$. Assuming this,
\begin{equation}
	\Box \calA_a^0 = [(\partial^2_x + \partial^2_y) \calA_0] k_a \equiv (\nabla_\perp^2 \calA_0 )k_a.
\end{equation}
Inspection of \eqref{xPortEM} thus shows that if $\nabla^2_\perp \calA_0 = 0$, it is consistent to set $\mathcal{A}^n_a = 0$ for all $n \geq 1$. The geometric-optics potential $a_a = \calA_0 e^{i \omega \varphi} k_a$ is then exact (and the exponential is superfluous in the sense that it may be absorbed into a redefinition of $\calA_0$). This potential is not pure gauge unless $\calA_0$ is independent of $x$ and $y$; more generally,
\begin{equation}
	\mathcal{F}^1_{ab} = - i k_{[a} \nabla_{b]} \calA_0
	\label{F1special}
\end{equation}
and $\mathcal{F}^n_{ab} = 0$ for all $n \geq 2$. Physically, these are exact solutions describing plane-fronted waves propagating in the $+z$ direction. The one non-vanishing field strength coefficient $\mathcal{F}_{ab}^1$ is essentially identical to a geometric-optics expression, except that it arises at one order higher and the polarization state is encoded in the gradient of $\calA_0$ instead of $e_+$ and $e_-$. More precisely, the coefficient \eqref{F1special} is equal to the zeroth-order coefficient $\mathcal{F}_{ab}^0$ which would be associated with the amplitude $\calA^0_a = \calA_0 (-i \nabla_a \ln \calA_0)$. 

Leaving aside this particular example, including a nonzero $\chi \calA_0 k_a$ term in $\calA^0_a$ is generically very similar to adding homogeneous solutions to the higher-order $\calA^n_a$. These terms essentially just reorder the 1-parameter family of solutions associated with the high-frequency approximation, and are therefore of minimal interest.

\subsection{Gravity}
\label{Sect:kPolgrav}

In a gravitational context, setting $e_\pm = 0$ in \eqref{eExpandgrav} results in the leading-order amplitude $\calA_{ab}^0 = \calA_0 k_{(a} \chi_{b)}$, where $k \cdot \nabla \chi_a = 0$. In these cases, $\mathcal{R}^0_{abcd} = 0$. To see how curvature perturbations may nevertheless arise at higher orders, consider the phase function $\varphi = t-z$ and the scalar amplitude $\calA_0 = \calA_0(\varphi,x,y)$ on a Minkowski background. Then the geometric-optics metric perturbation $h_{ab} = \calA_0 k_{(a} \chi_{b)} e^{i \omega \varphi}$ is exact (at least in linearized general relativity) if 
\begin{equation}
	\nabla^2_\perp \calA_0 = 0, \qquad \nabla_a \chi_b =0, \qquad (k \cdot \chi) \nabla_a \calA_0 = (\chi \cdot \nabla \calA_0) k_a. 
\end{equation}
The last of these conditions arises from \eqref{hGauge} and is satisfied in the Kerr-Schild case where $\chi_a \propto k_a$ (although it may also hold more generally). Regardless, these constraints and \eqref{RcoeffDef} imply that the subleading curvature perturbation is given by
\begin{equation}
	\mathcal{R}^1_{abcd} = -\frac{i}{2} k_{[a} ( \chi_{b]} \nabla_{[c} \calA_0 + \nabla_{b]} \calA_0 \chi_{[c}) k_{d]}. 
	\label{R1simp}
\end{equation}
A nonzero expression can also arise at one order beyond this,
\begin{equation}
	\mathcal{R}^2_{abcd} = \frac{1}{2} \big[ (k_{[a} \nabla_{b]}) \chi_{[c} \nabla_{d]} + (\chi_{[a} \nabla_{b]}) k_{[c} \nabla_{d]} \big] \calA_0,
	\label{R2simp}
\end{equation}
but $\mathcal{R}^n_{abcd} = 0$ for all $n \geq 3$.

The $\mathcal{R}^1_{abcd}$ given by \eqref{R1simp} has the same form as the geometric-optics curvature $\mathcal{R}^0_{abcd}$ which would be associated with the amplitude $\calA^0_{ab} = \calA_0 ( - i \chi_{(a} \nabla_{b)} \ln \calA_0)$. The $\mathcal{R}^2_{abcd}$ given by \eqref{R2simp} is different. It cannot necessarily be written in the geometric-optics form $k_{[a} (\ldots)_{b][c} k_{d]}$. However, significant simplifications arise for Kerr-Schild perturbations in which $\chi_a \propto k_a$: There is no loss of generality in making such a proportionality into an equality, in which case $\mathcal{R}^1_{abcd} = 0$ and $\mathcal{R}^2_{abcd}$ is equal to the geometric-optics $\mathcal{R}^0_{abcd}$ generated by the polarization tensor $- \calA_0^{-1} \nabla_a \nabla_b \calA_0$. Polarization states in these cases are thus controlled by second derivatives of $\calA_0$ instead of by $e_\pm$.

Kerr-Schild metric perturbations with the given form are in fact exact solutions not only to the linearized Einstein equation, but also to its fully-nonlinear counterpart. This is a consequence of the general result that Einstein's equation linearizes for metric perturbations proportional to $k_a k_b$, where $k_a$ is any null geodesic vector field; see \cite{ExactSolns, Xanthopoulos1978, HarteVines}. The specific Kerr-Schild solutions considered in this example are referred to as $pp$-waves: plane-fronted waves with parallel rays. Gravitational plane waves arise as special cases in which $\calA_0(\varphi,x,y)$ is quadratic in $x$ and $y$. Plane-fronted waves which are instead described more conventionally using the $e_\pm$ coefficients in \eqref{eExpandgrav} are related by a linear gauge transformation (which is not bounded as $\omega \to \infty$). However, the Kerr-Schild representation satisfies the exact Einstein equation while the one employing $e_\pm$ does not. This suggests the possibility of formulating a high-frequency approximation in which Kerr-Schild structures play a more central role, as enforced by, e.g., a gauge condition which replaces the Lorenz constraint \eqref{LorenzGrav}. While such a modification is likely to complicate various aspects of the linearized theory, it may bring considerable advantages in nonlinear contexts.

\ack

I thank Yi-Zen Chu, Sam Dolan, and Justin Vines for valuable discussions.

\section*{References}

\bibliographystyle{iopart-num}
\bibliography{refsLens}

\end{document}